\documentclass[iop]{emulateapj}
\providecommand{\mb}{$\Delta m_{15}(B)$}

\providecommand{\mu}{$\Delta m_{15}(U)$}

\begin{document}
%%%%%%%%%% Title & Author %%%%%%%%%%%%%%%%%%%%%%%%%%%%%%%%%%%%%%%%%%%%%%%%%%%%%%%
\title{Ultraviolet Observations of Super-Chandrasekhar Mass \\
Type I\lowercase{a} Supernova Candidates with Swift UVOT}

\author{Peter~J.~Brown\altaffilmark{1},
Paul Kuin\altaffilmark{2}, 
Richard Scalzo\altaffilmark{3}, 
Michael Smitka\altaffilmark{1}, \\
Massimiliano de Pasquale\altaffilmark{2}, 
Stephen Holland\altaffilmark{4}, 
Kevin Krisciunas\altaffilmark{1},\\
Peter Milne\altaffilmark{5}, 
Lifan Wang\altaffilmark{1}} %, 
\altaffiltext{1}{George P. and Cynthia Woods Mitchell Institute for Fundamental Physics \& Astronomy, 
Texas A. \& M. University, Department of Physics and Astronomy, 
4242 TAMU, College Station, TX 77843, USA; 
pbrown@physics.tamu.edu}            
\altaffiltext{2}{Mullard Space Science Laboratory, University College London, Holmbury St. Mary, Dorking Surrey, RH5 6NT, UK}

\altaffiltext{3}{Research School of Astronomy and Astrophysics, The Australian National University, Mount Stromlo Observatory, Cotter Road, Weston Creek ACT 2611, Australia}            

\altaffiltext{4}{Space Telescope Science Center
3700 San Martin Dr., 
Baltimore, MD 21218, USA}

\altaffiltext{5}{Steward Observatory, University of Arizona, Tucson, AZ 85719, USA}

%%%%%%%%%% Abstract %%%%%%%%%%%%%%%%%%%%%%%%%%%%%%%%%%%%%%%%%%%%%%%%%%%%%%%%%%%%
\begin{abstract}

Among Type Ia supernovae (SNe~Ia) exist a class of overluminous objects whose ejecta mass is inferred to be larger than the canonical Chandrasekhar mass.  We present and discuss the UV/optical photometric light curves, colors, absolute magnitudes, and spectra of three candidate Super-Chandrasekhar mass SNe--2009dc, 2011aa, and 2012dn--observed with the {\sl Swift} Ultraviolet/Optical Telescope.   The light curves are at the broad end for SNe Ia, with the light curves of SN~2011aa being amongst the broadest ever observed.  We find all three to have very blue colors which may provide a means of excluding these overluminous SNe from cosmological analysis, though there is some overlap with the bluest of ``normal'' SNe Ia.  All three are overluminous in their UV absolute magnitudes compared to normal and broad SNe Ia, but SNe 2011aa and 2012dn are not optically overluminous compared to normal SNe Ia. 
The integrated luminosity curves of SNe 2011aa and 2012dn in the UVOT range (1600-6000 \AA) are only half as bright as SN~2009dc, implying a smaller $^{56}$Ni yield. 
While not enough to strongly affect the bolometric flux, the early time mid-UV flux makes a significant contribution at early times.  The strong spectral features in the mid-UV spectra of SNe 2009dc and 2012dn suggest a higher temperature and lower opacity to be the cause of the UV excess rather than a hot, smooth blackbody from shock interaction.  Further work is needed to determine the ejecta and $^{56}$Ni masses of SNe 2011aa and 2012dn and fully explain their high UV luminosities.

\end{abstract}

%%%%%%%%%% Keywords %%%%%%%%%%%%%%%%%%%%%%%%%%%%%%%%%%%%%%%%%%%%%%%%%%%%%%%%%%%%
%%%%%%%%%%%%%%%%%%%%%%%%%%%%%%%%%%%%%%%%%%%%%%%%%%%%%%%%%%%%%%%%%%%%%%%%%%%%%%%%%
%\clearpage

\section{Candidate Super-Chandrasekhar mass Type I\lowercase{a} Supernovae  \label{intro}}

Type Ia Supernovae (SNe Ia) are important cosmological probes that first revealed the accelerating expansion of the universe \citep{Riess_etal_1998,Perlmutter_etal_1999}.
The cosmological results rely on the normal SNe Ia whose brightness correlates with their light curve shapes and colors \citep{Phillips_etal_1999,Riess_etal_1996_mlcs,Goldhaber_etal_2001}, allowing them to be used as standardizable candles.  Observations of similar but peculiar objects are useful for understanding the nature of the progenitor systems and the physics of the explosion, particularly how they might differ between objects.  It is also important to understand objects which may be found in cosmological samples but do not follow the relationships between the luminosity and the light curve shape.

The similar peak luminosities of SNe Ia suggested explosions of similar mass and energy.  The widely-held theory is that a SN Ia results from the thermonuclear disruption of a Carbon-Oxygen white dwarf (CO-WD) as it approaches the Chandrasekhar limit.  This could be due to accretion from a non-degenerate companion (also called the single degenerate scenario; \citealp{Whelan_Iben_1973}) or the disruption of a WD companion (also called the double degenerate scenario; \citealp{Webbink_1984, Iben_Tutukov_1984}).  The nature of an SN Ia progenitor as a C-O WD (and admittedly for a single case) has only recently been confirmed by very early time observations of SN~2011fe \citep{Nugent_etal_2011,Bloom_etal_2012}.  
The WD mass at explosion might not need approach the Chandrasekhar limit, as helium shell detonations can trigger a core detonation in sub-Chandrasekhar mass progenitors \citep{Woosley_Weaver_1994,Fink_etal_2010,Kromer_etal_2010,Woosley_Kasen_2011}.

The nature of the companion remains unknown, and recent results suggest that SNe Ia may result from both single degenerate and double degnerate systems.  Early observations of many SNe Ia do not show the interaction expected \citep{Kasen_2010} if the SN explosion were to interact with a red giant (RG) companion \citep{Hayden_etal_2010_shock,Bianco_etal_2011, Mo_etal_2011, Brown_etal_2012_shock}.  X-ray limits also rule out red giants due to the lack of shock interaction \citep{Russell_Immler_2012}.  Pre-explosion, multi-wavelength, and extremely early observations of SN~2011fe  rule out a RG \citep{Nugent_etal_2011,Li_etal_2011,Horesh_etal_2012,Margutti_etal_2012} and even a main sequence (MS) companion \citep{Bloom_etal_2012,Brown_etal_2012_11fe} for that object.  
Searches for the leftover companion in SNR 0509-67.5 rule out a non-degenerate companion \citep{Schaefer_Pagnotta_2012}.  On the other hand, high resolution spectroscopy of nearby SNe has found a preference for blue shifted sodium absorption in about 20-25\% of SNe Ia \citep{Sternberg_etal_2011,Foley_etal_2012_prog,Maguire_etal_2013} and even variable absorption \citep{Patat_etal_2007,Simon_etal_2009} suggestive of a local CSM wind from a non-degenerate companion.  PTF11kx observations showed signatures of a recurrent nova progenitor in a single degenerate system \citep{Dilday_etal_2012}.  Thus, multiple channels might be required to create the explosions classified as SNe Ia.

The idea that the accreting progenitor explodes as it approaches the Chandrasekhar mass has been challenged by a class of SNe that appear spectroscopically similar to SNe Ia but are overluminous for their light curve shape. Detailed modeling of the light curves appears to require more than a Chandrasekhar mass of ejected material.  
SN~2003fg was the first discovered \citep{Howell_etal_2006} with SNe 2006gz \citep{Hicken_etal_2007}, 2007if \citep{Scalzo_etal_2010, Yuan_etal_2010} and 2009dc \citep{Yamanaka_etal_2009,Tanaka_etal_2010,Silverman_etal_2011,Taubenberger_etal_2011,Kamiya_etal_2012,Hachinger_etal_2012} showing similarities.  \citet{Scalzo_etal_2012} discovered five additional, similar objects in SN Factory observations, though only one was conclusively above the Chandrasekhar limit.  Association with this subclass is sometimes based on spectroscopic similarity to others of the class, to a high inferred luminosity, or to actually modeling the light curve and determining a high ejecta mass.  Variations exist amongst candidates of his subclass, which is not surprising given our limited understanding of their origin and relationship to normal SNe Ia.  \citet{Maeda_Iwamoto_2009} highlight the observational differences between SNe 2003fg and 2006gz, two probable super-Chandarasekhar mass candidates.

The most common means of estimating the mass from SNe Ia comes from the application of ``Arnett's Law'' \citep{Arnett_1982,Branch_1992}.  At maximum light the luminosity output is approximately   equal to the instantaneous rate of energy release from radioactive decay.  Thus the peak bolometric luminosity is proportional to the mass of $^{56}$Ni synthesized in the explosion.  
The $^{56}$Ni can also be estimated from the late light curve \citep{Silverman_etal_2011} or nebular spectra  \citep{Mazzali_etal_1997}.  The total mass can be estimated based on energetics using the observed luminosities and expansion velocities and assumptions on the density profile (e.g. \citealp{Scalzo_etal_2012}).  The mass can also be estimated by constructing models of various masses and explosion scenarios and comparing to the observed light curves \citep{Kamiya_etal_2012} and spectra \citep{Mazzali_etal_1997,Hachinger_etal_2012}.

Not all of the luminosity necessarily comes from radioactive decay.  Excess luminosity could also come from circumstellar interaction \citep{Taubenberger_etal_2013} or result from asymmetric explosions viewed at a favorable angle \citep{Hillebrandt_etal_2007}.  Asymmetric explosions cannot explain the brightest of SC SNe, and spectropolarimetry of SN~2009dc implies no large scale asymmetries in the plane of the sky \citealp{Tanaka_etal_2010}).  \citet{Maeda_etal_2009} find that the late time observations of SN~2006gz require less radioactive Ni than suggested from peak optical observations, drawing into question the overluminous nature of the event.  They suggest that the luminosity is overestimated due to an over-correction for extinction. 

SC SNe are hot, high-energy explosions, so ultraviolet (UV) coverage is important to better measure the total luminosity and determine its origin, in particular whether it originates from shocks or simply a hot photosphere.
 The Ultraviolet/Optical Telescope (UVOT; \citealp{Roming_etal_2005}) on the Swift satellite \citep{Gehrels_etal_2004} presents an excellent opportunity to obtain unique, early-time UV data.
This paper will focus on three objects: SN~2009dc--a well-studied member of the Super-Chandrasekhar mass SN class--and SNe 2011aa and 2012dn which share some characteristics. 
We will refer to these candidate super-Chandrasekhar SNe Ia as SC SNe below, though a firm mass determination will require more data and is beyond the scope of this work.  
Comparisons will focus on the differences and similarities between SN~2009dc and the less studied SNe 2011aa and 2012dn, and the differences of these three SC SNe compared to other SNe Ia.  
In Section \ref{obs} we discuss these three SC SNe and present UV/optical photometry and spectra from UVOT.  In Section \ref{results} we compare the colors, absolute magnitudes, spectra, and integrated luminosities, comparing SNe 2011aa and 2012dn to 2009dc and the three to a larger sample of ``normal'' SNe Ia.   In Section \ref{discussion} we discuss the results and summarize.
%%%%%%%%%%%% 
 
\section{Swift Observations of Candidate Super-Chandrasekhar Mass SN\lowercase{e} } \label{obs}

\subsection{SN2009dc}

SN~2009dc was discovered by \citet{Puckett_etal_2009dc} on 2009 April 9.31 (all dates UT).  \citet{Marion_etal_2009dc} reported spectroscopic similarities to SC SNe on April 22.  Swift observations began  April 25.5.  Swift/UVOT photometry has been published by \citet{Silverman_etal_2011} and also referred to by \citet{Taubenberger_etal_2011}.  An epoch of UV grism spectroscopy was performed May 1.0 (4.9 days after the time of maximum light in the B-band).  
SN~2009dc has been extensively studied \citep{Yamanaka_etal_2009,Tanaka_etal_2010,Silverman_etal_2011,Taubenberger_etal_2011} including theoretical modeling of the light curves \citep{Kamiya_etal_2012} and spectra \citep{Hachinger_etal_2012}.
Assuming that its luminosity is powered by radioactive decay, SN~2009dc likely had a $^{56}$Ni yield between 1.2 and 1.8 M$_\odot$ depending on the assumed extinction (though \citealp{Silverman_etal_2011} also calculate a $^{56}$Ni mass of 3.7  M$_\odot$ for their largest plausible reddening).

SN~2009dc exploded outside of UGC 10064 toward the disrupted companion UGC 10063 (see \citealp{Taubenberger_etal_2011} and \citealp{Khan_etal_2011} for further discussion of the host environment).  The redshift of UGC 10064 is 0.021391 $\pm$ 0.000070 \citep{Falco_etal_1999}.   The foreground galactic extinction along the line of sight is A$_V$=0.191 \citep{Schlafly_Finkbeiner_2011}.

\subsection{SN2011aa}

SN~2011aa was discovered by \citet{Puckett_etal_2011} on 2011 February 6.3.  It was also independently discovered by MASTER on 2011 February 13.54 \citep{Kudelina_etal_2011}.  From optical spectra taken February 8.9 it was spectroscopically identified as a young SN Ia by \citet{Gurugubelli_etal_2011} who found a best match to the normal SN Ia 1998aq a week before maximum light.   Observations with the Swift spacecraft began on Feb 11.6 and continued for 16 epochs of UV and optical photometry (every other day around maximum light and then more spread out at later times).  One epoch of spectroscopy with the UVOT's UV grism was performed on February 28.0 (8.1 days after maximum light in the B-band), but overlap with a field star significantly contaminates the spectrum.   \citet{Kamiya_2012} found photometric similarities between SN~2011aa optical observations and SC SNe Ia models.  

SN~2011aa is located at the intersection of two galaxies designated UGC3906 at a redshift of 0.012355 +/- 0.000087 \citep{RC3}.  The foreground galactic extinction along the line of sight is A$_V$=0.078 \citep{Schlafly_Finkbeiner_2011}.

\begin{figure} 
\resizebox{6.8cm}{!}{\includegraphics*{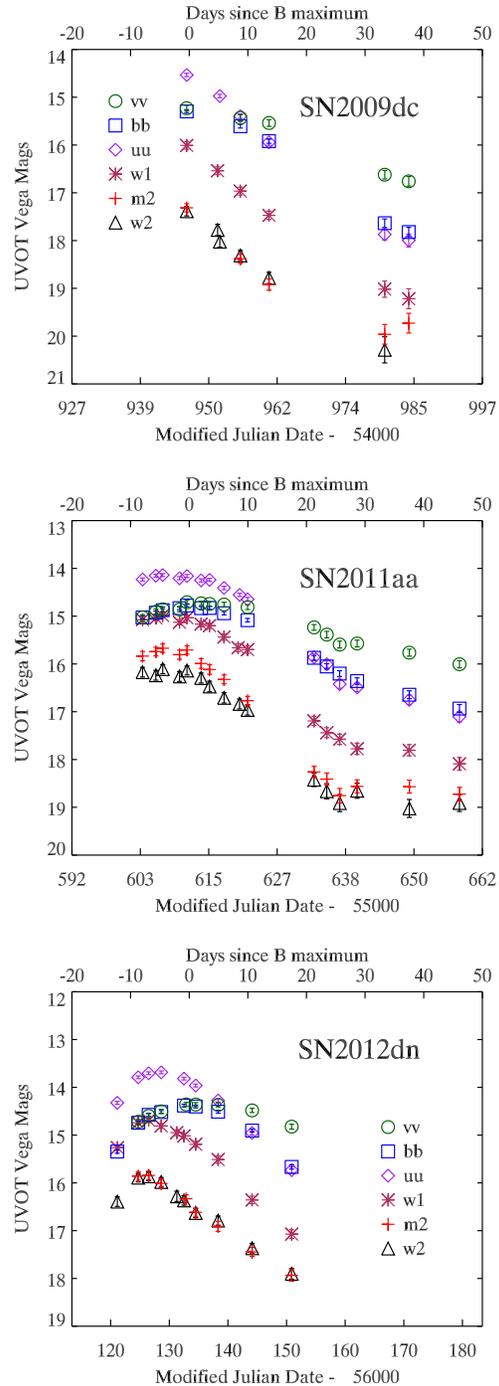}   }
\caption[Results]
        {UVOT light curves of SNe~2009dc, 2011aa, and 2012dn.  The same axis ranges are used for all SNe for a fair comparison.
 } \label{fig_lightcurves}    
\end{figure}

%\clearpage 
\begin{figure} 
%\resizebox{8.8cm}{!}{\includegraphics*{f2_color.eps}   }
\plotone{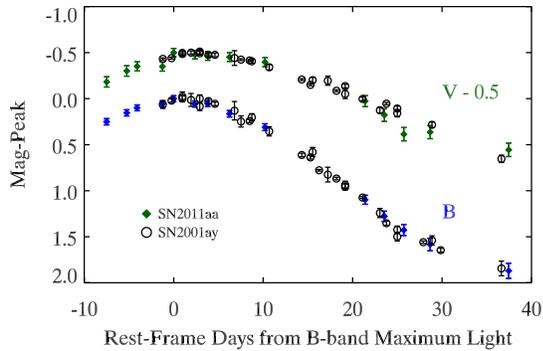}  
\caption[Results]
        {The b and v light curves of SN~2011aa compared to ground-based B and V of SN~2001ay from \citet{Krisciunas_etal_2011}.  The curves are shifted along the x-axis to the time of maximum light in the B band.  The curves are shifted vertically to the peakmagnitude in the respective filter, with the v curves shifted by an additional 0.5 mag for clarity.  The light curves of SN~2011aa are nearly identical to those of ``the most slowly declining type Ia supernova'' \citep{Krisciunas_etal_2011}.  
 } \label{fig_11ay01ay}    
\end{figure}

\subsection{SN2012dn}

SN~2012dn was discovered by \citet{Bock_etal_2012} on 2012 July 8.5. \citet{Parrent_Howell_2012dn} spectroscopically classified it as a SN Ia before maximum light from a spectrum obtained July 10.2.  They noted strong CII absorption and similarities to three SNe Ia described as SC SNe Ia.  \citet{Copin_etal_2012} also noticed similarities to SC SNe Ia spectra.  Swift observations began July 13.1.  One epoch of UV grism spectroscopy was obtained on July 22.6 (2.2 days before maximum light in the B band).  We also use an optical spectrum obtained by the South African Large Telescope (SALT) on July 24 from Parrent et al. (2013, in prep).
The UVOT and SALT spectra were combined by normalizing to the same B-band magnitude and splicing together at 4750 \AA.

SN~2012dn is located in the galaxy ESO 462-G016 at a redshift of 0.010187 $\pm$ 0.000020 \citep{Theureau_etal_1998}.  The foreground galactic extinction along the line of sight is A$_V$=0.167 \citep{Schlafly_Finkbeiner_2011}.

\subsection{Data Reduction}
The Swift/UVOT observations used the following six broadband filters with the corresponding central wavelengths \citep{Poole_etal_2008}: uvw2 (1928 \AA), uvm2 (2246 \AA), uvw1 (2600 \AA), u (3465 \AA), b (4392 \AA), and v (5468 \AA). Those filters are sometimes referred to as w2, m2, w1, uu, bb, and vv, respectively.
Swift/UVOT data were analyzed following the methods of \citet{Poole_etal_2008} and \citet{Brown_etal_2009} but incorporating the revised UV zeropoints and time-dependent sensitivity from \citet{Breeveld_etal_2011}.  The photometry is given in Table \ref{table_photometry}.  The light curves of the three SC SNe are displayed in Figure \ref{fig_lightcurves}.  The UVOT data for SN~2009dc were originally published in \citet{Silverman_etal_2011} and here we rereduce the data with the new zeropoints, sensitivity corrections, and subtraction of the underlying galaxy flux. 
The difference is small -- typically less than 0.05 mag.
The photometry for SN~2011aa also includes galaxy subtraction, so the late time flattening in the UV filters appears to be real.  SN~2012dn does not have galaxy template images, but the amount of galaxy contamination is likely small.  
The UVOT b and v bands are similar to Johnson $B$ and $V$ while the Swift u-band is extends to much shorter wavelengths than Johnson $U$ or Sloan $u'$ (and does not suffer from atmospheric attenuation), of particular importance for SNe such as these with different UV spectral shapes than normal SNe.
While we have obtained photometry in six bands, for simplicity we will focus on three filters with which to measure colors and absolute magnitudes.  We use uvm2 for the mid-UV (or MUV), uvw1 for the near-UV (or NUV), and the v-band for the optical.

The UVOT grism data was extracted and calibrated using the default parameters of the UVOTPY package (Kuin et al. 2014, in preparation).\footnotetext{www.mssl.ucl.ac.uk/www\_astro/uvot} The nominal wavelength accuracy is 20 \AA~ and the flux calibration is accurate to about 10\%.  Individual exposures were extracted and the spectra combined using a weighted mean.  

\subsection{Comparison Type Ia Supernovae }

For comparison, we use previously published photometry (updated to the latest calibration as described above) from spectroscopically normal SNe Ia with \mb $<$ 1.4 with detections in all three UV filters \citep{Brown_etal_2010,Brown_etal_2012_shock}.  For SN~2011fe we use spectrophotometry of SN~2011fe using the spectra from \citet{Pereira_etal_2013} due to the  UVOT data's optical saturation near peak \citep{Brown_etal_2012_11fe}.  Further comparisons are made with SNe spectroscopically similar to SN~2002cx (SNe 2005hk and 2012Z) and SN~1991T (SNe 2007S, 2007cq, and 2011dn).  The photometry from 2011dn is presented here for the first time while SN~2012Z will be presented in Stritzinger et al. (2014, in preparation) and the others were previously published in \citet{Brown_etal_2009}.

\section{Results}\label{results} 

\subsection{Light Curves \label{curves}}

The light curves of SNe 2009dc, 2011aa and 2012dn are shown in Figure \ref{fig_lightcurves}.  Swift observations of SN~2009dc began near maximum light so the curves monotonically fade, but all appear qualitatively very similar, including the crossing points of the different filters.  The evolution of SN~2011aa, though, is much slower.  We show in Figure \ref{fig_11ay01ay} that it has a similar decay rate in b and v as SN~2001ay, ``the most slowly declining type Ia supernova'' \citep{Krisciunas_etal_2011}.  The UV light curves are also broader than those of normal SNe Ia, most of which have very similar post-maximum light curve shapes in the NUV \citep{Immler_etal_2006,Milne_etal_2010}.  We parameterize the light curves of SNe 2011aa and 2012dn by their peak magnitudes and by $\Delta $m$_{15}$, the number of magnitudes they fade in the 15 days following maximum light in that same band.  The peak is determined by stretching a template light curve to the data between 5 days before and 5 days after maximum light. $\Delta $m$_{15}$ is determined by stretching a template light curve to the data between 2 days before and 15 days after maximum light to the data and interpolating from the stretched template.  The UV templates (with uvw1 template also being used for u band due to its similar light curve shape) come from SN~2011fe \citep{Brown_etal_2012_11fe} and B and V from MLCS2k2 \citep{Jha_etal_2007}.  We note that for very broad SNe such as some of these, the measurement of light curve parameters depend heavily on how the fitting and determination of the peak time is done.  The light curve parameters are tabulated in Table \ref{table_parameters}.  We also list the difference in time between when the SN peaks in the b band compared to the other filters.  The difference in peak times between b and uvw1 are much larger than the normal SNe analyzed by \citet{Milne_etal_2010} with a mean of 2.22 days and the largest being 3.7.  SN~2009dc is excluded since observations began near the optical peak, while the UV was already fading.

The b-band light curves are found to peak at MJD 54947.1 (2009 April 26.1), 55611.9 (2011 February 19.9), and 56132.8 (2012 July 24.8) for SNe 2009dc, 2011aa, and 2012dn, respectively.  These values are used as the reference epochs for the light curves and spectra displayed.

\begin{deluxetable}{llrrrrr} 
\tablecaption{UVOT Photometry \label{table_photometry}} 
\tablehead{\colhead{SN} & \colhead{Filter} & \colhead{MJD} & \colhead{Mag} & \colhead{M\_Err} &  \colhead{Rate} & \colhead{R\_Err}   }  
\startdata 

SN2012dn   & UVW2     & 56121.1333 &  16.393 &   0.100 &   2.482 &   0.230 \\
SN2012dn   & UVM2     & 56124.7118 &  15.857 &   0.089 &   2.495 &   0.204 \\
SN2012dn   & UVW1     & 56121.1419 &  15.262 &   0.056 &   7.434 &   0.383 \\
SN2012dn   & U        & 56121.1303 &  14.324 &   0.031 &  40.397 &   1.152 \\
SN2012dn   & B        & 56121.1313 &  15.344 &   0.036 &  32.096 &   1.059 \\
SN2012dn   & V        & 56124.7077 &  14.720 &   0.048 &  18.529 &   0.822 \\

\enddata 
\tablecomments{ The full table is available in the electronic version.  The photometry will also be available from the Swift SN website http://people.physics.tamu.edu/pbrown/SwiftSN/swift\_sn.html. } 
\end{deluxetable}

\begin{deluxetable}{lrr} 
\tablecaption{Light Curve Parameters \label{table_parameters}} 
\tablehead{\colhead{Parameters} & \colhead{SN2011aa} & \colhead{SN2012dn}   }  
\startdata

  m$_{w2}$(peak) &    16.15 $ \pm $    0.04 &   15.86 $ \pm $    0.04 \\ 
 $\Delta $m$_{15}$(w2)  &    0.97 $ \pm $    0.10 &    1.03 $ \pm $    0.06 \\ 
 t$_{max}$(w2)-t$_{max}$(b) &   -4.4 $ \pm $    1.4~~ &   -7.1 $ \pm $    0.5~~ \\ 
 
  m$_{m2}$(peak) &    15.70 $ \pm $    0.04 &   15.84 $ \pm $    0.15 \\ 
 $\Delta $m$_{15}$(m2)  &    1.15 $ \pm $    0.08 &    1.02 $ \pm $    0.05 \\ 
 t$_{max}$(m2)-t$_{max}$(b) &   -3.7 $ \pm $    1.2~ &   -7.3 $ \pm $    1.6~ \\ 
 
  m$_{w1}$(peak) &    15.01 $ \pm $    0.02 &   14.71 $ \pm $    0.02 \\ 
 $\Delta $m$_{15}$(w1)  &    0.78 $ \pm $    0.06 &    1.21 $ \pm $    0.05 \\ 
 t$_{max}$(w1)-t$_{max}$(b) &   -4.9 $ \pm $    0.8~ &   -7.0 $ \pm $    0.4~ \\ 
 
  m$_{u}$(peak) &    14.14 $ \pm $    0.01 &   13.69 $ \pm $    0.01 \\ 
 $\Delta $m$_{15}$(u)  &    0.67 $ \pm $    0.02 &    1.14 $ \pm $    0.03 \\ 
 t$_{max}$(u)-t$_{max}$(b) &   -4.3 $ \pm $    0.5~ &   -5.0 $ \pm $    0.3~ \\ 
 
  m$_{b}$(peak) &    14.80 $ \pm $    0.01 &   14.38 $ \pm $    0.07 \\ 
 $\Delta $m$_{15}$(b)  &    0.59 $ \pm $    0.07 &     1.08 $ \pm $     0.03 \\ 
 
  m$_{v}$(peak) &    14.73 $ \pm $    0.02 &   14.36 $ \pm $    0.10 \\ 
 $\Delta $m$_{15}$(v)  &    0.30 $ \pm $    0.07 &     0.44 $ \pm $     0.04 \\ 
 t$_{max}$(v)-t$_{max}$(b) &    2.4 $ \pm $    0.7~ &    0.6 $ \pm $    1.8~ \\ 
\enddata 
%\tablecomments{  } 
\end{deluxetable} 

%%%%%%%%;;;;;;;;;;;;;;;;;;;;;;;;;;;;   C O L O R S   ;;;;;;;;;;;;;;;;;;;;;;

\begin{figure*} 
\plottwo{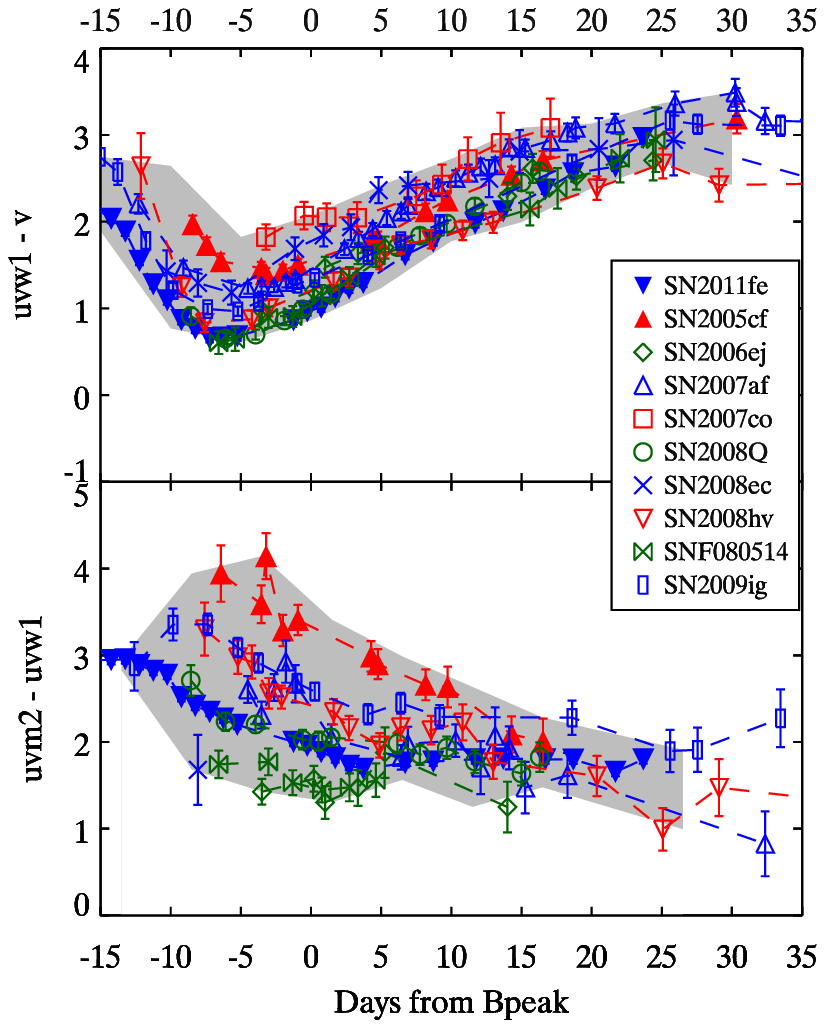} {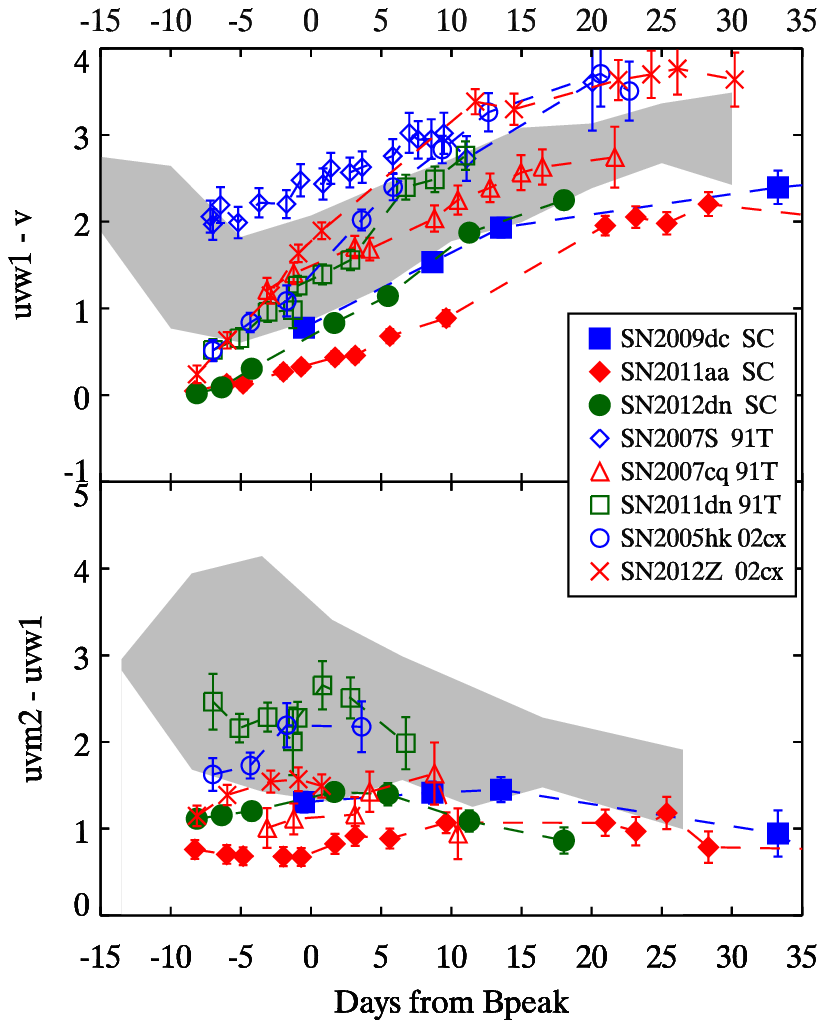} 
\caption[Results]
        {Left: uvw1-v and uvm2-uvw1 colors of normal SNe Ia observed by Swift with shaded region showing their range of colors.
The identification of individual color curves is not as important as the range of colors exhibited by ``normal'' SNe Ia.  
Right: uvw1-v and uvm2-uvw1 colors of the three SC SNe Ia showed with respect to the shaded region of normal SN Ia colors.  For comparison, SNe similar to SNe 1991T and 2002cx are also plotted.  SC SNe Ia are distinctly bluer than the normal SNe Ia, but some of the 1991T-like and 2002cx-like SNe can be just as blue.
 } \label{fig_colors}    
\end{figure*}

\begin{figure*} 
\plottwo{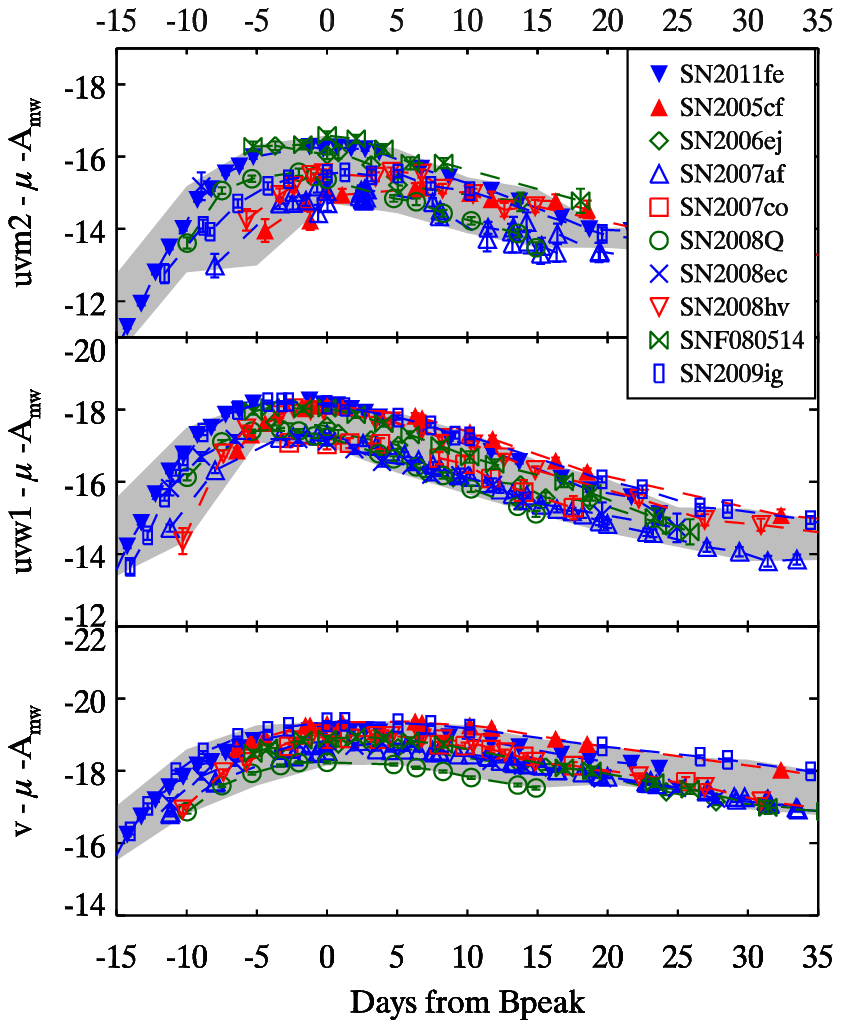} {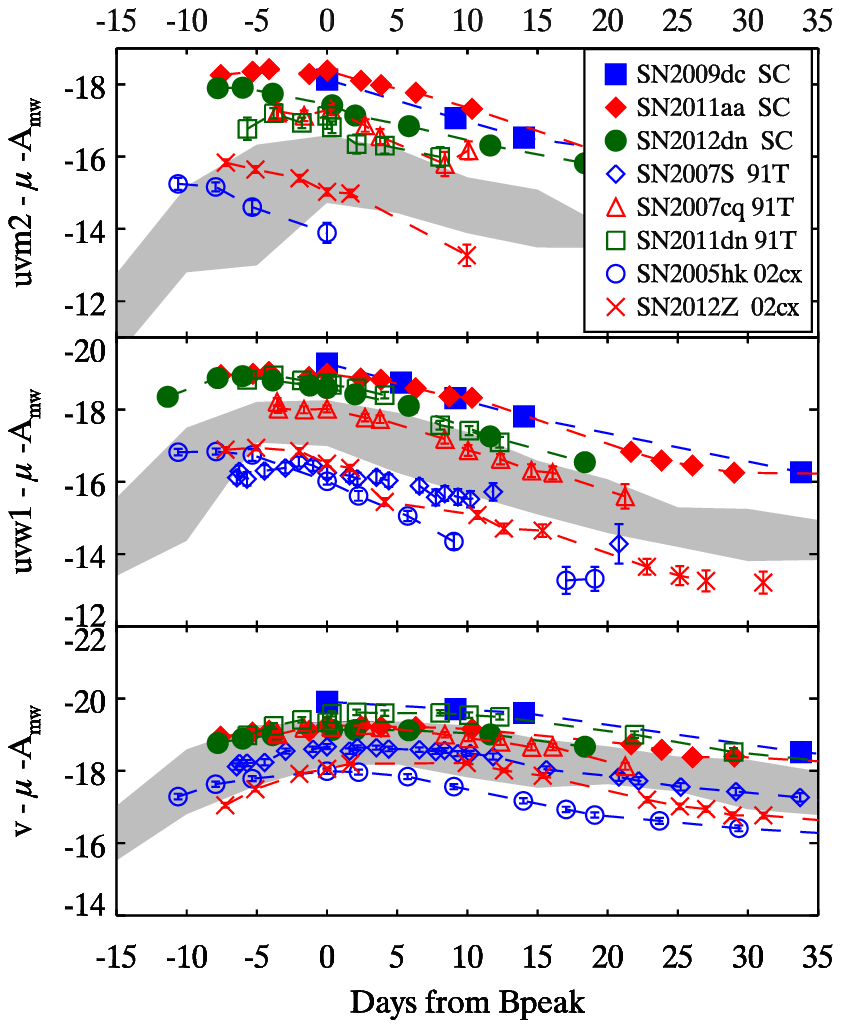} 
\caption[Results]
        {Left: Absolute magnitudes (correcting for distance modulus and MW extinction)  of normal SNe Ia observed by Swift with shaded region showing their range.
The SNe are labeled the same as in the left panel of Figure \ref{fig_colors}.  The curves of individual SNe are not as important as the range of absolute magnitudes at various epochs.  
Right: Absolute magnitudes (correcting for distance modulus and MW extinction)  of the three SC SNe Ia showed with respect to the shaded region of normal SN Ia absolute magnitudes.  For comparison, SNe similar to SNe 1991T and 2002cx are also plotted.  SC and 1991T-like SNe Ia are at the bright end of the normal distribution in the optical and distinctly brighter in the UV.  
The 2002cx-like SNe are at the faint end of the normal distribution in the optical but become relatively brighter (and peak much earlier) in the mid-UV.
 } \label{fig_abscurves}    
\end{figure*}

\subsection{Colors \label{colors}}

Figure \ref{fig_colors} shows the color evolution in uvm2-uvw1 and uvw1-v of the three SC SNe Ia compared to spectroscopically normal Swift SNe with $\Delta $m$_{15}$(B) $< $1.4.  The left panel shows that the uvw1-v colors of normal SNe evolve from red to blue, reaching a  minimum color a few days before the optical maximum and then becoming redder again.  The uvm2-uvw1 colors of SNe Ia have a large dispersion and tend to become slowly bluer.  The range of normal SN colors is compared to our SC sample in the right panel.  SN~2009dc, which was first observed near the optical maximum, is at the blue end of both colors but not extremely so.  SN~2012dn has similar colors at similar epochs but was also observed at earlier epochs.  At those pre-maximum epochs SN~2012dn was bluer than the normal SNe.  SN~2011aa had similar premaximum colors to SN~2012dn but did not redden as quickly as the others due to the slower light curve evolution shown above.

\citet{Milne_etal_2013} suggest that the spread in the NUV colors of normal SNe Ia can be viewed as two separate subclasses.  In addition to their bluer colors, the NUV-blue subclass also shares a spectroscopic trait with SC SNe in showing CII in their optical spectra \citet{Thomas_etal_2011,Milne_etal_2013}.  
SNe 2003fg \citep{Howell_etal_2006}, 2006gz \citep{Hicken_etal_2007}, 2007if \citep{Yuan_etal_2010,Scalzo_etal_2010}, 2009dc \citep{Taubenberger_etal_2011, Silverman_etal_2011}, and 2012dn \citep{Parrent_Howell_2012dn} 
all noted CII, often very strong.  Two slow-decliners that aren't considered SC candidates, 2001ay and 2009ig, may have had weak CII features \citep{Krisciunas_etal_2011,Foley_etal_2012_09ig}.  

While SN~2009dc is actually included in the NUV-blue subclass (as the bluest member) in \citet{Milne_etal_2013}, the early phase observations of SNe 2011aa and 2012dn presented here show that the SC SNe are much bluer than normal SNe Ia at earlier times.  SC SNe may not have the early red to blue evolution of normal SNe Ia, or it happens earlier than ten days before optical maximum.  Their slower evolution, on the other hand, might make the time of optical maximum a poor reference point for giving physical meaning to their behavior compared to normal SNe Ia.  Nevertheless, it is clear that the SC SNe Ia extend the diversity in the UV more than that already found \citep{Brown_etal_2009,Wang_etal_2012,Milne_etal_2013}.  
Early UV observations appear to be a way of photometrically separating SC SNe from normal SNe.  This could be quantified as the magnitude or timing of the minimum color (i.e. the color at its bluest epoch) or the time difference between maximum light in the UV and optical bands.

Two additional classes of SNe also warrant further comparison.  Spectroscopic similarity to SN~1991T was used as a follow up criterion to discover new SC candidates by \citet{Scalzo_etal_2012}.  SNe similar to SN~2002cx (also called SNe Iax; \citealp{Foley_etal_2013_Iax}) also show hot, highly-ionized photospheres.  We are not making a physical connection between the groups, but they warrant further comparison because of how their similar physical conditions have a strong effect on their UV flux and because of possible confusion in spectroscopic classification \citep{Foley_etal_2013_Iax}.  Several examples of each are displayed in the right hand panel of Figure \ref{fig_colors}.  
Similar to the SC SNe, 1991T-like and 2002cx-like SNe showed a monotonic reddening in uvw1-v from the onset of Swift observations.  With the exception of SN~2007S (whose optical colors suggest reddening from the host galaxy; \citealp{Brown_etal_2010}), all could have been as blue (in uvw1-v) as the SC SNe if they were observed early enough but the colors became redder at a much faster rate than the SC SNe.  In the uvm2-uvw1 color, one of each class had a comparable color.  The 91T-like SN~2007cq was classified by \citet{Milne_etal_2013} as a MUV-blue, because it follows the NUV-red group in uvw1-v but is relatively blue in uvm2-uvw1.  Thus multi-epoch multi-wavelength UV photometry reveals complicated similarities and differences amongst SNe of different subclasses and within the same subclass.  Further observations of members of these classes, including UV spectroscopy and even earlier UV photometry, will help explain the physical origins of the UV flux.

\begin{figure} 
\resizebox{8.8cm}{!}{\includegraphics*{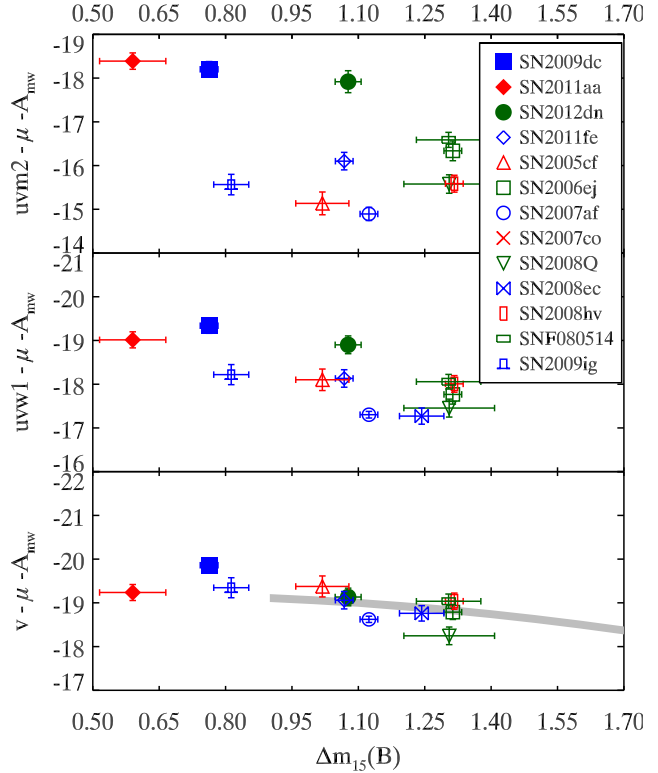}   }
\caption[Results]
        {Peak Absolute magnitudes (correcting for distance modulus and MW extinction)  of normal and candidate SC SNe Ia observed by Swift.  The absolute magnitude v. \mb~relation of \citet{Phillips_etal_1999} is plotted as a grey band with a width corresponding to the 1 $\sigma$ uncertainty.  The Swift sample of normal SNe Ia is consistent with the relation but has a large scatter (due primarily in the optical to distance uncertainties; see \citealp{Brown_etal_2010}).  Of the three SC SNe Ia observed by Swift, all three are overluminous in the UV but only SN~2009dc is overluminous in the optical.  The y-axis is the same in all three panels to show how the scatter in absolute magnitudes increases to shorter wavelengths \citep{Brown_etal_2010} as does the separation in brightness between the SC SNe and the normal SNe Ia.  
 } \label{fig_dmb_abs}    
\end{figure}

\begin{figure} 
\plotone{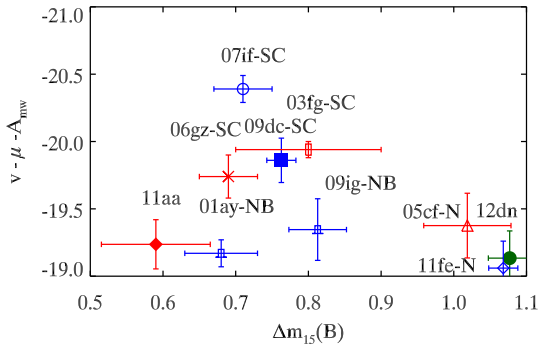} 
\caption[Results]
        {Peak Absolute magnitudes in the V band (correcting for distance modulus and MW extinction)  of normal and candidate SC SNe Ia with \mb $<$ 1.1 observed by Swift with additional broad SNe observed from ground-based facilities.  
SN~2009dc has a comparable V-band brightness as the other probable Super Chandrasekhar-mass SNe (marked with 'SC').
The two new SC SN candidates discussed here have optical absolute magnitudes comparable to the normal but broad ('NB')
 SNe 2001ay and 2009ig and the very normal SNe 2005cf and 2011fe (marked with a 'N'). } \label{fig_dmb_vabs}    
\end{figure} 

\subsection{Absolute Magnitudes \label{absmags}}

Since one common characteristic of the strongest SC SN candidates is their high luminosity, we now examine the absolute magnitudes of SNe 2009dc, 2011aa, and 2012dn.  
As in \citet{Brown_etal_2010}, most distance moduli are computed from the host galaxy recessional velocity, corrected for local velocity flows \citep{Mould_etal_2000}, and a Hubble constant of 72 km/s/Mpc \citep{Freedman_etal_2001}.  Distances from Cephieds, the Tully-Fisher relation, or surface brightness fluctuations are used when available, as listed in \citet{Brown_etal_2010} and \citet{Brown_etal_2012_11fe}.
For SNe 2009dc, 2011aa, and 2012dn the adopted Hubble flow distances are 34.94 $ \pm $  0.16, 33.894 $ \pm $  0.18, and 33.324 $ \pm $   0.20, respectively.
The SNe in this sample are at relatively low redshifts (mostly with recessional velocities less than 6000 km/s), so thermal velocities of the galaxies can add a significant dispersion to distances calculated assuming a Hubble flow.  Thus the scatter of absolute magnitudes in the optical is much larger than found for larger samples of SNe (see \citealp{Brown_etal_2010} for more details on this sample).  
\citet{Maeda_etal_2009} suggest that the extinction and thus the luminosity of SN~2006gz could be overestimated.  To avoid such overcorrections, we do not correct any of the SNe Ia for host extinction.  We do correct for line sight extinction in the Milky Way (MW) by converting the \citet{Schlafly_Finkbeiner_2011} V-band extinction from NED to an E(B-V) reddening (by dividing by 3.1) and then multiplying by the extinction coefficients calculated for the UV-optical spectrum of SN~1992A \citep{Brown_etal_2010}.  

Figure \ref{fig_abscurves} shows the absolute magnitudes in the optical (v band), NUV (uvw1), and MUV (uvm2).  While SC SNe Ia are brighter than most (but not all; see below) in the optical, they are almost one magnitude brighter than the brightest in the NUV and about two magnitudes brighter in the MUV.  The light curve shapes are similar in shape, but SN~2012dn fades a little faster while SNe 2009dc and 2011aa remain brighter than normal SNe Ia for a month after peak.  
For comparison, SNe similar to SNe 1991T and 2002cx are also plotted.  SC and 1991T-like SNe Ia are at the bright end of the normal distribution in the optical and distinctly brighter in the UV.  
The 2002cx-like SNe are at the faint end of the normal distribution in the optical but become relatively brighter (and peak much earlier) at shorter wavelengths.

Figure \ref{fig_dmb_abs} shows the peak absolute magnitudes compared to \mb~ for the normal and SC SNe Ia.  The SC SNe candidates all have broad optical light curves (i.e. low values of \mb) but not uniquely broad.   In the UV, all three are noticeably brighter. The peak optical luminosities of SNe 2011aa and 2012dn are comparable to those of the  normal SNe Ia.  SN~2009dc is significantly brighter in the optical.  Figure \ref{fig_dmb_vabs} zooms in on the absolute v-band magnitudes of the broad SNe, including ground based observations of other broad SNe.  SN~2009dc lies clearly amongst the other SC SNe while SNe 2011aa and 2012dn have v-band absolute magnitudes consistent with the other SNe Ia. 
The absolute magnitudes, especially in the UV, are very sensitive to the extinction.  One concern for the analysis of SN~2006gz based on the luminosity is that it is very sensitive to the assumed extinction -- SN~2006gz could be fainter if there is less host extinction or if the extinction coefficient is smaller.  In multiple analyses SN~2009dc is bright even if no host galaxy extinction is assumed but could be even brighter.  In this plot we have assumed no host dust extinction, yet they could be extinguished by dust in the host galaxy, and thus intrinsically brighter.  
The extremely blue colors would suggest that the host reddening is minimal, but the intrinsic colors of these objects are not actually known.  A larger sample is needed to determine observationally what the range of colors might be and how blue the unreddened color could be.  
The degeneracy between reddening and luminosity means that the SNe 2011aa and 2012dn could have low reddening and be optically underluminous compared to SN~2009dc.  Alternatively, significant reddening would mean they are intrinsically bluer and even more overluminous in the UV.  Either way, these three SNe show similarities but are not identical.

\begin{figure*} 
\resizebox{15cm}{!}{\includegraphics*{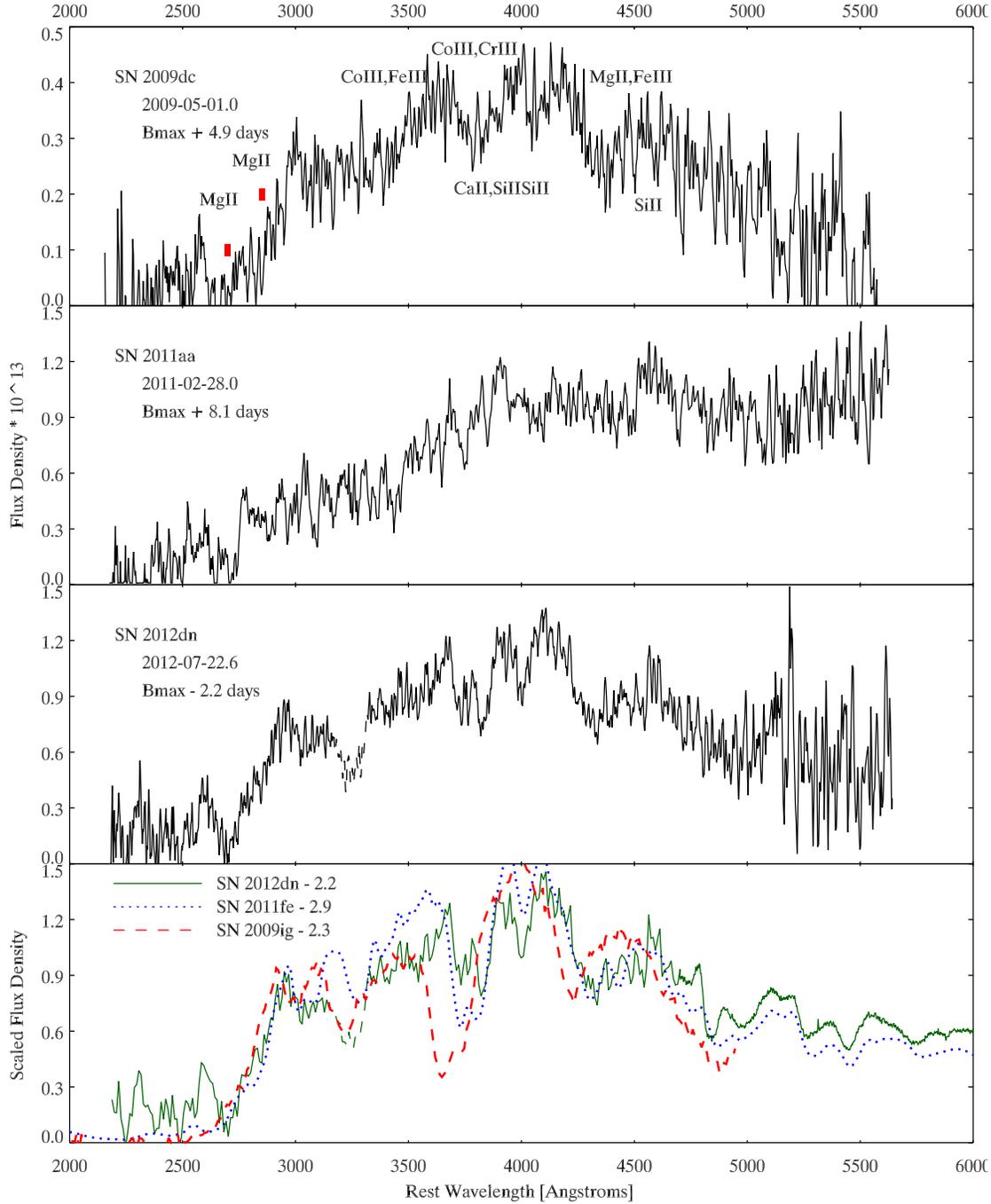}   }
\caption[Results]
        {Top: UVOT grism spectrum of SN~2009dc. The position of NUV MgII lines are marked (for a velocity of 9000 km/s).  The other line identifications are from \citet{Hachinger_etal_2012}.
	Top Middle:  UVOT grism spectrum of SN~2011aa.   There is significant contamination from an overlapping stellar spectrum.  This spectrum is shown here for completeness but not utilized further.
	Bottom Middle:  UVOT grism spectrum of SN~2012dn. The absorption region between 3250 and 3400, set apart by a dashed line, is due to a readout streak from a bright source in the background region.
Bottom: Combined spectrum of SN~2012dn compared to normal SNe Ia 2011fe and broad 2009ig.  For comparison purposes, all spectra are normalized to have the same B band magnitude.  The NUV continuum of SN~2012dn is not significantly different than the NUV-blue SN~2011fe, while SN~2009ig has broader absorption features.    
Shortward of 2700 \AA, SN~2012dn has a clear MUV excess composed of strong features rather than the smooth continuum with diluted features expected if the excess luminosity were due to a hot blackbody spectrum from shock interaction.
 } \label{fig_uvotspectra}    
\end{figure*}

\subsection{Spectral Comparisons \label{spectra_results}}

Figure \ref{fig_uvotspectra} shows the UVOT grism spectra of SNe 2009dc, 2011aa, and 2012dn.  The signal to noise ratio is much lower than usual for optical SN spectroscopy, but comparable to other Swift/UVOT spectra \citep{Bufano_etal_2009,Foley_etal_2012_09ig}.  SN~2011aa was contaminated by an overlapping stellar spectrum, but SNe 2009dc and 2012dn exhibit similar continuum shapes and features.  Absorption from MgII appears in the NUV, while shortward of that 
the spectrum is blanketed by overlapping lines of iron-peak elements.
The bottom panel of Figure \ref{fig_uvotspectra} compares the combined UVOT-SALT spectrum of SN2012dn to an HST UV/optical spectrum of SN~2011fe taken 2011 September 7 (2.9 days before maximum light in the B band; \citealp{Mazzali_etal_2013}) and a Swift/UVOT grism spectrum of the broad but normal SN~2009ig \citep{Foley_etal_2012_09ig} taken 2009 September 3.7 (2.3 days before maximum light in the B band).  All three spectra have been normalized to the same b-band magnitude to compare the relative flux in the UV.  

SN~2011fe, classified as a NUV-blue SN \citep{Milne_Brown_2012,Milne_etal_2013}, and SN~1009ig are not dissimilar to SN~2012dn above 4000 \AA.  The CaII H\&K lines of SN~2009ig are very broad and deep \citep{Foley_etal_2012_09ig,Marion_etal_2013}, reducing its NUV flux.  In the MUV, SNe~2009ig and 2011fe have a much lower flux and a smoother pseudocontinuum.   While we do not want to overinterpret the grism spectrum by studying individual features at this time, the strong undulations in the MUV of SNe 2009dc and 2012dn suggest a lower opacity (and thus gaps in the line blanketing), rather than a hot blackbody from a shock interaction, as the source of the increased UV luminosity.

%\clearpage 
\begin{figure} 
\resizebox{8.8cm}{!}{\includegraphics*{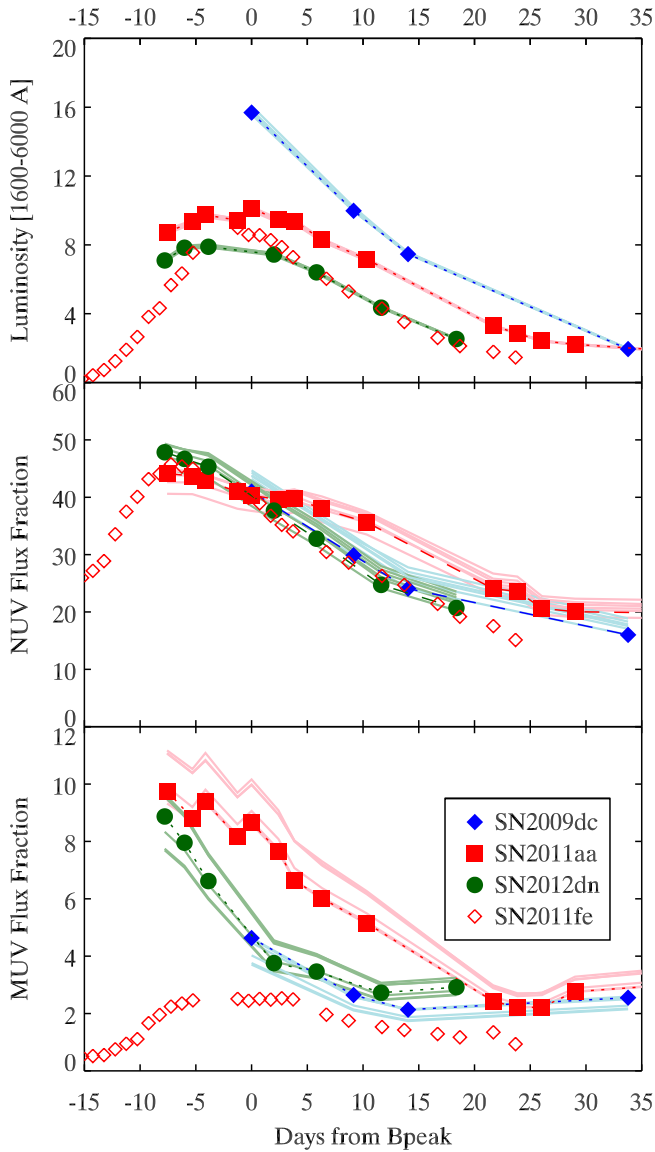}   }
\caption[Results]
        { Top: Integrated flux from 1600 to 6000 \AA.  SN~2009dc is about twice as bright as SNe 2011aa and 2012dn whcih have about the same brightness as SN~2011fe.  Faint lines in all three plots represent flux reconstructions using different spectral templates.  The difference is negligible in the integrated luminosity so the lines overlap.
Middle: Fractions of the 1600-6000 \AA~integrated flux in the NUV (2800 - 4000 \AA) region for the SC SNe Ia compared to the NUV-blue normal SN~2011fe.  The NUV fractions for normal SNe Ia peak between 30 (for NUV-red SNe Ia) and 40\% (For NUV-blue) with the SC SNe Ia all near 40\%.
Bottom: Fractions of the 1600-6000 \AA~integrated flux in the MUV (1600 - 2800 \AA) region for the SC SNe Ia compared to the NUV-blue normal SN~2011fe. SNe 2012dn and 2011aa have 10 and 9\%, respectively, of their flux in the MUV in their earliest observation.  By B-band maximum light, the fraction for SN~2012dn has dropped down to 4\%, only modestly above SN~2011fe.  SN~2011aa has a significantly larger fraction of its flux (compared to the others) for at least ten days after the B-band maximum.  
 } \label{fig_flux}    
\end{figure}

\subsection{Integrated Luminosity \label{absmags}}

To determine how much flux is observed, we need to convert from the observed magnitudes.   Flux conversion factors are very spectrum dependent in the UV \citep{Brown_etal_2010}, differing by source type and phase for objects (like SNe) with time-variable spectra.  Simpler SEDs based on the photometry have a problem reproducing the multi-filter photometry.  To estimate the flux contributions of different wavelength regions, we use the combined UV/optical spectrum we have for SN~2012dn and warp it to match the photometry as follows.  First, the spectrum was smoothed with a running average over 10 \AA. The spectrum is extrapolated beyond the UVOT filter range using the mean of the spectrum in the shortest 50 \AA~(between 2200 and 2250 \AA).  Then the whole spectrum was scaled by a constant value to match the observed b-band magnitude at that epoch.  A warping function is created from linear segments from 1500 \AA~ to 8100 \AA~ (just beyond the UVOT bounds) with pivot points near where the UVOT filter curves intersect each other (after normalizing by the integral of the effective area of the curve to deweight the broader filters).  These points are at  2030, 2460, 3050, 3870, and 4960 \AA.  These seven points are iteratively adjusted to minimize the magnitude differences between the observed photometry and that of the warped spectrum.  This method better reproduces the spectral shape than converting the observed photometry to independent flux density points and simply connecting the dots (Brown et al. 2014, in preparation).  The SN2012dn spectrum is used for SNe 2012dn and 2011aa.  For SN~2009dc we combine the UVOT grism spectrum with a comparable epoch spectrum from \citet{Taubenberger_etal_2011} obtained from the WISEREP database\footnote{http://www.weizmann.ac.il/astrophysics/wiserep/} \citep{Yaron_Gal-Yam_2012}.  

From these warped spectra we calculate the amount of flux coming from the full UVOT range (1600 - 6000 \AA) and three regions -- MUV (1600 - 2800 \AA), NUV (2800 - 4000 \AA), and optical (4000 - 6000 \AA) at each epoch with photometry in all six UVOT filters. 
The top panel of Figure \ref{fig_flux} shows the integrated flux of the three SC SNe compared to a direct integration of the UV/optical spectra of \citet{Pereira_etal_2013} for the normal NUV-blue SN~2011fe.  Despite their bright UV luminosity, SNe 2011aa and 2012dn which have about the same integrated luminosity as the normal SN~2011fe.  SN~2009dc is about twice as bright.  
 The evolution of the MUV and NUV flux fractions (compared to the total 1600 - 6000 \AA~flux) are displayed in the middle and lower panel of Figure \ref{fig_flux}.  The NUV fractions for the SC SNe and SN2011fe all peak between 44 and 48\%.
SNe 2012dn and 2011aa have 10 and 9\%, respectively, of their flux in the MUV in their earliest observation.  By B-band maximum light, the fraction for SN~2012dn has dropped down to 4\%, only modestly above SN~2011fe.  SN~2011aa has a significantly larger fraction of its flux (compared to the others) for at least ten days after the B-band maximum.  As a simple check on the effect of the template spectrum we perform the same color matching for all three SNe using the SNe 2009dc and 2012dn spectra,  UV/optical spectra of SN~2011fe from \citet{Pereira_etal_2013} near maximum light and 24 days after maximum, the near maximum spectrum of the SN Ia 1992A from \citet{Kirshner_etal_1993} and a spectrum of Vega from \citet{Bohlin_Gilliland_2004}.  The total flux changes by only a few percent and the MUV fraction changes by up to 15\% (in a relative sense), indicating that the luminosity measurement is dependent on, but not dominated by, the spectral template inputed.

The integrated luminosity curves allow us to compare in a relative sense the bolometric luminosity of SNe 2011aa and 2012dn to the well-studied SN~2009dc, and thus the inferred $^{56}$Ni mass.  Since the colors of SNe 2011aa and 2012dn are similar to or bluer in color than SN 2009dc, we assume for now that they do not suffer significantly more dust extinction than SN~2009dc and that the same fractions of the bolometric luminosities lie outside of our 1600-6000 \AA~range for all three SNe.  We also assume the rise time is similar for the three SNe and that the ratio of the bolometric luminosity to the radioactive luminosity is the same.  Under these (many) assumptions, the mass of $^{56}$Ni is proportional to the integrated luminosity L$_{1600-6000}$.  
For a range of host galaxy reddening values, \citet{Silverman_etal_2011} determined a 
$^{56}$Ni between 1.2 and 3.7 $M_{\odot}$, with a most likely value of 1.7 $\pm$ 0.4 $M_{\odot}$.
Since SNe 2011aa and 2012dn have about half the integrated luminosity, their $^{56}$Ni would likely be around 0.9 $M_{\odot}$.  While smaller than SN~2009dc, it is close to the the amount of $^{56}$Ni produced (0.92 $M_{\odot}$) in a Chandrasekhar-mass detonation where the entire mass is converted into iron group elements \citep{Khokhlov_etal_1993}.  One could also make the comparison with SN~2011fe.  Because of its smaller luminosity and shorter rise time (16.58 days for SN~2011fe compared to at least 21.1 days for SN~2009dc), the  $^{56}$Ni mass is estimated to be 0.53 $M_{\odot}$ \citep{Pereira_etal_2013}.  If SNe 2011aa and 2012dn were assumed to have similar rise times and radiative efficiencies as SN~2011fe, the  $^{56}$Ni masses would also be similar.  So the estimate relies in part on assumptions about an unobserved property--the rise time.  The observed properties of SNe 2011aa and 2012dn are more similar to SN~2009dc than SN~2011fe, but this highlights the need for a better understanding these objects and limiting the assumptions that must be made.  Further analysis is needed to determine more accurately the  $^{56}$Ni mass required and what progenitor/explosion scenarios might result in these observables.

%%%%%%%%%%%%%%%%%%%%%%%%%%%%%%%%%%%%%%%%%%%%%%%%%%%% 

%%%%%%%%%%%%
%%%%%%%%%%%%%%%%%%%%%%%%%%%%%%%%%%%%%%%%%%%%%%%%%%%%%%%%%%%%%%%%%%%%%%
\section{Discussion and Summary} \label{discussion}

One suggestion for the increased luminosity in SC SNe is shock interaction \citep{Fryer_etal_2010, Blinnikov_Sorokina_2010, Taubenberger_etal_2011,
Hachinger_etal_2012}. 
\citet{Scalzo_etal_2012} suggest UV observations as a means to probe the influence of shock interactions on the early luminosity.  \citet{Fryer_etal_2010} performed numerical calculations of the spectra and simulated UVOT light curves resulting from a double-degnerate SN Ia exploding within a shell of unaccreted material.  While our candidate SC SNe Ia have peak luminosities comparable to those studied by \citet{Fryer_etal_2010}, the light curves shapes are much different.  The light curve shapes can vary based on the amount and spatial distribution of the surrounding material, but the smoothness of the UV light curves and their qualitative similarity to the optical light curves suggest a photospheric origin.  

A photospheric origin for the emission is supported by the UV spectra of SNe 2009dc and 2012dn, which show stronger features in the MUV (below 2700 \AA) than seen in normal SNe Ia, while the flux from a hot shock would be relatively smooth and would dilute the photospheric features \citep{Hamuy_etal_2003}.  On the other hand, the optical features are also much stronger than for SN~2007if, for which the top lighting of a shock was invoked as one explanation for its diluted features and high luminosity \citep{Scalzo_etal_2010}. The UV spectra of SNe 2009dc and 2012dn do not allow a smooth blackbody source for the excess flux.  Such a spectrum might be expected from a high temperature shock with a hydrogen-rich circumstellar medium, as was used to explain the diluted features of SN~2002ic (\citealp{Hamuy_etal_2003}, see also \citealp{Branch_etal_2000}).  A structured spectrum with emission and absorption, due to reprocessing of the shock emission or originating from a different composition, cannot be excluded.  \citet{Hachinger_etal_2012} found adding a spectrum of the Ibn SN~2006jc to their theoretical SN Ia spectrum gave reasonable matches to the observed spectra of SN~2009dc.  Higher quality UV spectra of Ibn and SC SNe are needed to perform similar tests in the UV.  
Nevertheless, photometric observations may already contain enough information to further constrain photometric (e.g. \citealp{Kamiya_etal_2012}) or spectroscopic modeling (e.g. \citealp{Hachinger_etal_2012}).

While the optical light curves  of SNe 2011aa and 2012dn are not dissimilar to normal SNe Ia (though extremely broad in the case of SN~2011aa), the NUV-optical and especially the MUV-NUV (or MUV-optical) colors are markedly different.  The rest-frame UV also peaks earlier for SC than normal SNe Ia.  Early rest-frame UV photometry might allow optically overluminous SNe such as SN~2009dc to be excluded from cosmological analysis.   
\citet{Scalzo_etal_2012} estimate the rate of SC SNe to be a few percent of all SNe Ia locally, but a bias could result from an evolutionary shift \citep{Taubenberger_etal_2011} if these are more common in the early universe than they are locally.  
\citet{Milne_etal_2013_z} show that the relative fractions of NUV-blue and NUV-red normal SNe Ia change with redshift.  
The origin of the UV diversity amongst normal and SC candidate SNe Ia may %be a way 
point to ways to reduce the dispersion at longer wavelengths and understand potential biases in SN Ia standardization at different epochs in the history of the universe.

Bolometric light curve comparisons between models and observations serve as an important diagnostic of allowed models and parameters.  The creation of bolometric light curves, however, especially the treatment of missing wavelength ranges, varies greatly.  Sometimes the NUV (or at least the ground based U band) is included, and the MUV may or may not be included. Often the UV portion of the flux is considered to be negligible (a reasonable assumption in some cases).  If it is included, it is often set at a constant percentage of the flux.  As shown here, there is also a lot of variation in the NUV and MUV flux fractions between various SNe Ia, and the fractions evolve quite significantly with time.   

The data given here will allow the bolometric light curves of these objects to be more accurately determined.  For example, the falling UV fraction means that inclusion of the UV flux will broaden the pre-maximum rise of the bolometric flux.  This could lead to a longer implied rise time if fit with a light curve template.  This longer rise time may not be accurate, however, if the stretched light curve template did not include the UV in its construction.    \citet{Kamiya_etal_2012} use multi-wavelength modeling to show the difference between a BVRI, UV-Optical-IR (UVOIR), and true bolometric light curve.  The distinction between these is important.  UV data will allow more constraints on the modelling.

While we have pushed the knowledge of the UV behavior for SC SNe Ia $\sim8$ days earlier, the very earliest epochs would also be important for looking for the effects of shock interaction with a non-degenerate companion \citep{Kasen_2010,Brown_etal_2012_shock} or differences in the UV-optical flux evolution at the earliest times \citep{Brown_etal_2012_11fe}.  As the UV-optical colors are still bluest at the first epochs observed, the bolometric contribution before then may be larger still and are in any case uncertain.  Higher quality UV spectra at the earliest possible epochs will better probe the mechanism responsible for the excess UV emission and how to account for it in mass determinations.

In summary, we have presented UV/optical photometry and spectroscopy for three SNe Ia, 2009dc, 2011aa and 2012dn, which have been suggested as candidate super-Chandrasekhar mass SNe Ia.  While their optical properties are not dissimilar to normal SNe Ia, they are significantly bluer and more luminous in the UV than normal SNe Ia, with MUV luminosities about a factor of $\sim10$ higher.  UV spectra of SNe~2009dc and 2012dn feature structure not expected for shock interaction, suggesting a photospheric origin of the excess UV luminosity.  The UV is shown to contribute significantly (but still smaller than the optical) to the bolometric luminosity, especially at early times.  The integrated luminosities of SNe 2011aa and 2012dn are much lower than 2009dc, however. This suggests a larger diversity in the class, if they are indeed in the same class, when considering UV and optical photometric and spectroscopic characteristics.  A more detailed study of these SNe is required to determine if they were above the Chandrasekhar mass.

%%%%%%%%%%%%%%%%%%%%%%%%%%%%%%%%%%%%%%%%%%%%%%%%%%%%%%%%%%%%%%%%%%%%%%%%%%%%%%%%%

\acknowledgements

P.J.B. is supported by the Mitchell Postdoctoral Fellowship and NSF grant AST-0708873.  
P.A.M acknowledges support from NASA ADAP grant NNX10AD58G.
We are grateful to J. Vinko and J. Parrent for sharing the spectra of SN~2012dn.  
We are also grateful to H. Marion and X. Wang for looking at optical spectra of SN~2011aa.  
This work made use of public data in the {\it Swift} data
archive and the NASA/IPAC Extragalactic Database (NED), which is
operated by the Jet Propulsion Laboratory, California Institute of
Technology, under contract with NASA.  

%%%%%%%%%%%%%%%%%%%%%%%%%%%%%%%%%%%%%%%%%%%%%%%%%%%%%%%%%%%%%%%%%%%%%%%%%%%%%%%%%

%%%%%%%%%%%%%%%%%%%%%%%%%%%%%%%%%%%%%%%%%%%%%%%%%%%%%%%%%%%%%%%%%%%%%%%%%%%%%%%%%

\bibliographystyle{apj}

\begin{thebibliography}{94}
\expandafter\ifx\csname natexlab\endcsname\relax\def\natexlab#1{#1}\fi

\bibitem[{{Arnett}(1982)}]{Arnett_1982}
{Arnett}, W.~D. 1982, \apj, 253, 785

\bibitem[{{Bianco} {et~al.}(2011){Bianco}, {Howell}, {Sullivan}, {Conley},
  {Kasen}, {Gonz{\'a}lez-Gait{\'a}n}, {Guy}, {Astier}, {Balland}, {Carlberg},
  {Fouchez}, {Fourmanoit}, {Hardin}, {Hook}, {Lidman}, {Pain},
  {Palanque-Delabrouille}, {Perlmutter}, {Perrett}, {Pritchet}, {Regnault},
  {Rich}, \& {Ruhlmann-Kleider}}]{Bianco_etal_2011}
{Bianco}, F.~B., {Howell}, D.~A., {Sullivan}, M., {et~al.} 2011, ApJ, 741, 20

\bibitem[{{Blinnikov} \& {Sorokina}(2010)}]{Blinnikov_Sorokina_2010}
{Blinnikov}, S.~I., \& {Sorokina}, E.~I. 2010, ArXiv e-prints

\bibitem[{{Bloom} {et~al.}(2012){Bloom}, {Kasen}, {Shen}, {Nugent}, {Butler},
  {Graham}, {Howell}, {Kolb}, {Holmes}, {Haswell}, {Burwitz}, {Rodriguez}, \&
  {Sullivan}}]{Bloom_etal_2012}
{Bloom}, J.~S., {Kasen}, D., {Shen}, K.~J., {et~al.} 2012, \apjl, 744, L17

\bibitem[{{Bock} {et~al.}(2012){Bock}, {Parrent}, \& {Howell}}]{Bock_etal_2012}
{Bock}, G., {Parrent}, J.~T., \& {Howell}, D.~A. 2012, Central Bureau
  Electronic Telegrams, 3174, 1

\bibitem[{{Bohlin} \& {Gilliland}(2004)}]{Bohlin_Gilliland_2004}
{Bohlin}, R.~C., \& {Gilliland}, R.~L. 2004, \aj, 127, 3508

\bibitem[{{Branch}(1992)}]{Branch_1992}
{Branch}, D. 1992, \apj, 392, 35

\bibitem[{{Branch} {et~al.}(2000){Branch}, {Jeffery}, {Blaylock}, \&
  {Hatano}}]{Branch_etal_2000}
{Branch}, D., {Jeffery}, D.~J., {Blaylock}, M., \& {Hatano}, K. 2000, \pasp,
  112, 217

\bibitem[{{Breeveld} {et~al.}(2011){Breeveld}, {Landsman}, {Holland}, {Roming},
  {Kuin}, \& {Page}}]{Breeveld_etal_2011}
{Breeveld}, A.~A., {Landsman}, W., {Holland}, S.~T., {et~al.} 2011, in American
  Institute of Physics Conference Series, Vol. 1358, American Institute of
  Physics Conference Series, ed. J.~E. {McEnery}, J.~L. {Racusin}, \&
  N.~{Gehrels}, 373--376

\bibitem[{{Brown} {et~al.}(2012{\natexlab{a}}){Brown}, {Dawson}, {Harris},
  {Olmstead}, {Milne}, \& {Roming}}]{Brown_etal_2012_shock}
{Brown}, P.~J., {Dawson}, K.~S., {Harris}, D.~W., {et~al.} 2012{\natexlab{a}},
  ApJ, 749, 18

\bibitem[{{Brown} {et~al.}(2009){Brown}, {Holland}, {Immler}, {Milne},
  {Roming}, {Gehrels}, {Nousek}, {Panagia}, {Still}, \& {Vanden
  Berk}}]{Brown_etal_2009}
{Brown}, P.~J., {Holland}, S.~T., {Immler}, S., {et~al.} 2009, \aj, 137, 4517

\bibitem[{{Brown} {et~al.}(2010){Brown}, {Roming}, {Milne}, {Bufano},
  {Ciardullo}, {Elias-Rosa}, {Filippenko}, {Foley}, {Gehrels}, {Gronwall},
  {Hicken}, {Holland}, {Hoversten}, {Immler}, {Kirshner}, {Li}, {Mazzali},
  {Phillips}, {Pritchard}, {Still}, {Turatto}, \& {Vanden
  Berk}}]{Brown_etal_2010}
{Brown}, P.~J., {Roming}, P.~W.~A., {Milne}, P., {et~al.} 2010, ApJ, 721, 1608

\bibitem[{{Brown} {et~al.}(2012{\natexlab{b}}){Brown}, {Dawson}, {de Pasquale},
  {Gronwall}, {Holland}, {Immler}, {Kuin}, {Mazzali}, {Milne}, {Oates}, \&
  {Siegel}}]{Brown_etal_2012_11fe}
{Brown}, P.~J., {Dawson}, K.~S., {de Pasquale}, M., {et~al.}
  2012{\natexlab{b}}, \apj, 753, 22

\bibitem[{{Bufano} {et~al.}(2009){Bufano}, {Immler}, {Turatto}, {Landsman},
  {Brown}, {Benetti}, {Cappellaro}, {Holland}, {Mazzali}, {Milne}, {Panagia},
  {Pian}, {Roming}, {Zampieri}, {Breeveld}, \& {Gehrels}}]{Bufano_etal_2009}
{Bufano}, F., {Immler}, S., {Turatto}, M., {et~al.} 2009, \apj, 700, 1456

\bibitem[{{Copin} {et~al.}(2012){Copin}, {Gangler}, {Pereira}, {Rigault},
  {Smadja}, {Aldering}, {Birchall}, {Childress}, {Fakhouri}, {Kim}, {Nordin},
  {Nugent}, {Perlmutter}, {Runge}, {Saunders}, {Suzuki}, {Thomas}, {Pecontal},
  {Buton}, {Feindt}, {Kerschhaggl}, {Kowalski}, {Benitez}, {Hillebrandt},
  {Kromer}, {Sasdelli}, {Sternberg}, {Taubenberger}, {Baugh}, {Chen},
  {Chotard}, {Wu}, {Tao}, {Fouchez}, {Tilquin}, {Hadjiyska}, {Rabinowitz},
  {Baltay}, {Ellman}, {McKinnon}, {Effron}, {Cellier-Holzem}, {Canto},
  {Antilogus}, {Bongard}, \& {Pain}}]{Copin_etal_2012}
{Copin}, Y., {Gangler}, E., {Pereira}, R., {et~al.} 2012, The Astronomer's
  Telegram, 4253, 1

\bibitem[{{de Vaucouleurs} {et~al.}(1991){de Vaucouleurs}, {de Vaucouleurs},
  {Corwin}, {Buta}, {Paturel}, \& {Fouqu{\'e}}}]{RC3}
{de Vaucouleurs}, G., {de Vaucouleurs}, A., {Corwin}, Jr., H.~G., {et~al.}
  1991, {Third Reference Catalogue of Bright Galaxies. Volume I: Explanations
  and references. Volume II: Data for galaxies between 0$^{h}$ and 12$^{h}$.
  Volume III: Data for galaxies between 12$^{h}$ and 24$^{h}$.}

\bibitem[{{Dilday} {et~al.}(2012){Dilday}, {Howell}, {Cenko}, {Silverman},
  {Nugent}, {Sullivan}, {Ben-Ami}, {Bildsten}, {Bolte}, {Endl}, {Filippenko},
  {Gnat}, {Horesh}, {Hsiao}, {Kasliwal}, {Kirkman}, {Maguire}, {Marcy},
  {Moore}, {Pan}, {Parrent}, {Podsiadlowski}, {Quimby}, {Sternberg}, {Suzuki},
  {Tytler}, {Xu}, {Bloom}, {Gal-Yam}, {Hook}, {Kulkarni}, {Law}, {Ofek},
  {Polishook}, \& {Poznanski}}]{Dilday_etal_2012}
{Dilday}, B., {Howell}, D.~A., {Cenko}, S.~B., {et~al.} 2012, Science, 337, 942

\bibitem[{{Falco} {et~al.}(1999){Falco}, {Kurtz}, {Geller}, {Huchra}, {Peters},
  {Berlind}, {Mink}, {Tokarz}, \& {Elwell}}]{Falco_etal_1999}
{Falco}, E.~E., {Kurtz}, M.~J., {Geller}, M.~J., {et~al.} 1999, \pasp, 111, 438

\bibitem[{{Fink} {et~al.}(2010){Fink}, {R{\"o}pke}, {Hillebrandt},
  {Seitenzahl}, {Sim}, \& {Kromer}}]{Fink_etal_2010}
{Fink}, M., {R{\"o}pke}, F.~K., {Hillebrandt}, W., {et~al.} 2010, \aap, 514,
  A53

\bibitem[{{Foley} {et~al.}(2012{\natexlab{a}}){Foley}, {Simon}, {Burns},
  {Gal-Yam}, {Hamuy}, {Kirshner}, {Morrell}, {Phillips}, {Shields}, \&
  {Sternberg}}]{Foley_etal_2012_prog}
{Foley}, R.~J., {Simon}, J.~D., {Burns}, C.~R., {et~al.} 2012{\natexlab{a}},
  ArXiv e-prints

\bibitem[{{Foley} {et~al.}(2012{\natexlab{b}}){Foley}, {Challis}, {Filippenko},
  {Ganeshalingam}, {Landsman}, {Li}, {Marion}, {Silverman}, {Beaton},
  {Bennert}, {Cenko}, {Childress}, {Guhathakurta}, {Jiang}, {Kalirai},
  {Kirshner}, {Stockton}, {Tollerud}, {Vink{\'o}}, {Wheeler}, \&
  {Woo}}]{Foley_etal_2012_09ig}
{Foley}, R.~J., {Challis}, P.~J., {Filippenko}, A.~V., {et~al.}
  2012{\natexlab{b}}, \apj, 744, 38

\bibitem[{{Foley} {et~al.}(2013){Foley}, {Challis}, {Chornock},
  {Ganeshalingam}, {Li}, {Marion}, {Morrell}, {Pignata}, {Stritzinger},
  {Silverman}, {Wang}, {Anderson}, {Filippenko}, {Freedman}, {Hamuy}, {Jha},
  {Kirshner}, {McCully}, {Persson}, {Phillips}, {Reichart}, \&
  {Soderberg}}]{Foley_etal_2013_Iax}
{Foley}, R.~J., {Challis}, P.~J., {Chornock}, R., {et~al.} 2013, \apj, 767, 57

\bibitem[{{Freedman} {et~al.}(2001){Freedman}, {Madore}, {Gibson}, {Ferrarese},
  {Kelson}, {Sakai}, {Mould}, {Kennicutt}, {Ford}, {Graham}, {Huchra},
  {Hughes}, {Illingworth}, {Macri}, \& {Stetson}}]{Freedman_etal_2001}
{Freedman}, W.~L., {Madore}, B.~F., {Gibson}, B.~K., {et~al.} 2001, ApJ, 553,
  47

\bibitem[{{Fryer} {et~al.}(2010){Fryer}, {Ruiter}, {Belczynski}, {Brown},
  {Bufano}, {Diehl}, {Fontes}, {Frey}, {Holland}, {Hungerford}, {Immler},
  {Mazzali}, {Meakin}, {Milne}, {Raskin}, \& {Timmes}}]{Fryer_etal_2010}
{Fryer}, C.~L., {Ruiter}, A.~J., {Belczynski}, K., {et~al.} 2010, \apj, 725,
  296

\bibitem[{{Ganeshalingam} {et~al.}(2011){Ganeshalingam}, {Li}, \&
  {Filippenko}}]{Mo_etal_2011}
{Ganeshalingam}, M., {Li}, W., \& {Filippenko}, A.~V. 2011, \mnras, 416, 2607

\bibitem[{{Gehrels} {et~al.}(2004){Gehrels}, {Chincarini}, {Giommi}, {Mason},
  {Nousek}, {Wells}, {White}, {Barthelmy}, {Burrows}, {Cominsky}, {Hurley},
  {Marshall}, {M{\'e}sz{\'a}ros}, {Roming}, {Angelini}, {Barbier}, {Belloni},
  {Campana}, {Caraveo}, {Chester}, {Citterio}, {Cline}, {Cropper}, {Cummings},
  {Dean}, {Feigelson}, {Fenimore}, {Frail}, {Fruchter}, {Garmire}, {Gendreau},
  {Ghisellini}, {Greiner}, {Hill}, {Hunsberger}, {Krimm}, {Kulkarni}, {Kumar},
  {Lebrun}, {Lloyd-Ronning}, {Markwardt}, {Mattson}, {Mushotzky}, {Norris},
  {Osborne}, {Paczynski}, {Palmer}, {Park}, {Parsons}, {Paul}, {Rees},
  {Reynolds}, {Rhoads}, {Sasseen}, {Schaefer}, {Short}, {Smale}, {Smith},
  {Stella}, {Tagliaferri}, {Takahashi}, {Tashiro}, {Townsley}, {Tueller},
  {Turner}, {Vietri}, {Voges}, {Ward}, {Willingale}, {Zerbi}, \&
  {Zhang}}]{Gehrels_etal_2004}
{Gehrels}, N., {Chincarini}, G., {Giommi}, P., {et~al.} 2004, ApJ, 611, 1005

\bibitem[{{Goldhaber} {et~al.}(2001){Goldhaber}, {Groom}, {Kim}, {Aldering},
  {Astier}, {Conley}, {Deustua}, {Ellis}, {Fabbro}, {Fruchter}, {Goobar},
  {Hook}, {Irwin}, {Kim}, {Knop}, {Lidman}, {McMahon}, {Nugent}, {Pain},
  {Panagia}, {Pennypacker}, {Perlmutter}, {Ruiz-Lapuente}, {Schaefer},
  {Walton}, \& {York}}]{Goldhaber_etal_2001}
{Goldhaber}, G., {Groom}, D.~E., {Kim}, A., {et~al.} 2001, ApJ, 558, 359

\bibitem[{{Gurugubelli} {et~al.}(2011){Gurugubelli}, {Sahu}, {Anupama}, {Anto},
  \& {Arora}}]{Gurugubelli_etal_2011}
{Gurugubelli}, K., {Sahu}, D.~K., {Anupama}, G.~C., {Anto}, P., \& {Arora}, S.
  2011, Central Bureau Electronic Telegrams, 2653, 3

\bibitem[{{Hachinger} {et~al.}(2012){Hachinger}, {Mazzali}, {Taubenberger},
  {Fink}, {Pakmor}, {Hillebrandt}, \& {Seitenzahl}}]{Hachinger_etal_2012}
{Hachinger}, S., {Mazzali}, P.~A., {Taubenberger}, S., {et~al.} 2012, ArXiv
  e-prints

\bibitem[{{Hamuy} {et~al.}(2003){Hamuy}, {Phillips}, {Suntzeff}, {Maza},
  {Gonz{\'a}lez}, {Roth}, {Krisciunas}, {Morrell}, {Green}, {Persson}, \&
  {McCarthy}}]{Hamuy_etal_2003}
{Hamuy}, M., {Phillips}, M.~M., {Suntzeff}, N.~B., {et~al.} 2003, Nature, 424,
  651

\bibitem[{{Hayden} {et~al.}(2010){Hayden}, {Garnavich}, {Kasen}, {Dilday},
  {Frieman}, {Jha}, {Lampeitl}, {Nichol}, {Sako}, {Schneider}, {Smith},
  {Sollerman}, \& {Wheeler}}]{Hayden_etal_2010_shock}
{Hayden}, B.~T., {Garnavich}, P.~M., {Kasen}, D., {et~al.} 2010, ApJ, 722, 1691

\bibitem[{{Hicken} {et~al.}(2007){Hicken}, {Garnavich}, {Prieto}, {Blondin},
  {DePoy}, {Kirshner}, \& {Parrent}}]{Hicken_etal_2007}
{Hicken}, M., {Garnavich}, P.~M., {Prieto}, J.~L., {et~al.} 2007, ApJL, 669,
  L17

\bibitem[{{Hillebrandt} {et~al.}(2007){Hillebrandt}, {Sim}, \&
  {R{\"o}pke}}]{Hillebrandt_etal_2007}
{Hillebrandt}, W., {Sim}, S.~A., \& {R{\"o}pke}, F.~K. 2007, \aap, 465, L17

\bibitem[{{Horesh} {et~al.}(2012){Horesh}, {Kulkarni}, {Fox}, {Carpenter},
  {Kasliwal}, {Ofek}, {Quimby}, {Gal-Yam}, {Cenko}, {de Bruyn}, {Kamble},
  {Wijers}, {van der Horst}, {Kouveliotou}, {Podsiadlowski}, {Sullivan},
  {Maguire}, {Howell}, {Nugent}, {Gehrels}, {Law}, {Poznanski}, \&
  {Shara}}]{Horesh_etal_2012}
{Horesh}, A., {Kulkarni}, S.~R., {Fox}, D.~B., {et~al.} 2012, \apj, 746, 21

\bibitem[{{Howell} {et~al.}(2006){Howell}, {Sullivan}, {Nugent}, {Ellis},
  {Conley}, {Le Borgne}, {Carlberg}, {Guy}, {Balam}, {Basa}, {Fouchez}, {Hook},
  {Hsiao}, {Neill}, {Pain}, {Perrett}, \& {Pritchet}}]{Howell_etal_2006}
{Howell}, D.~A., {Sullivan}, M., {Nugent}, P.~E., {et~al.} 2006, \nat, 443, 308

\bibitem[{{Iben} \& {Tutukov}(1984)}]{Iben_Tutukov_1984}
{Iben}, Jr., I., \& {Tutukov}, A.~V. 1984, \apjs, 54, 335

\bibitem[{{Immler} {et~al.}(2006){Immler}, {Brown}, {Milne}, {The}, {Petre},
  {Gehrels}, {Burrows}, {Nousek}, {Williams}, {Pian}, {Mazzali}, {Nomoto},
  {Chevalier}, {Mangano}, {Holland}, {Roming}, {Greiner}, \&
  {Pooley}}]{Immler_etal_2006}
{Immler}, S., {Brown}, P.~J., {Milne}, P., {et~al.} 2006, ApJL, 648, L119

\bibitem[{{Jha} {et~al.}(2007){Jha}, {Riess}, \& {Kirshner}}]{Jha_etal_2007}
{Jha}, S., {Riess}, A.~G., \& {Kirshner}, R.~P. 2007, ApJ, 659, 122

\bibitem[{Kamiya(2012)}]{Kamiya_2012}
Kamiya, Y. 2012, Light Curve Model for SN 2011aa, poster presented at
  Supernovae Illuminating the Universe: from Individuals to Populations,
  September 10--14, Garching bei Muenchen, Germany

\bibitem[{{Kamiya} {et~al.}(2012){Kamiya}, {Tanaka}, {Nomoto}, {Blinnikov},
  {Sorokina}, \& {Suzuki}}]{Kamiya_etal_2012}
{Kamiya}, Y., {Tanaka}, M., {Nomoto}, K., {et~al.} 2012, \apj, 756, 191

\bibitem[{{Kasen}(2010)}]{Kasen_2010}
{Kasen}, D. 2010, ApJ, 708, 1025

\bibitem[{{Khan} {et~al.}(2011){Khan}, {Stanek}, {Stoll}, \&
  {Prieto}}]{Khan_etal_2011}
{Khan}, R., {Stanek}, K.~Z., {Stoll}, R., \& {Prieto}, J.~L. 2011, \apjl, 737,
  L24

\bibitem[{{Khokhlov} {et~al.}(1993){Khokhlov}, {Mueller}, \&
  {Hoeflich}}]{Khokhlov_etal_1993}
{Khokhlov}, A., {Mueller}, E., \& {Hoeflich}, P. 1993, \aap, 270, 223

\bibitem[{{Kirshner} {et~al.}(1993){Kirshner}, {Jeffery}, {Leibundgut},
  {Challis}, {Sonneborn}, {Phillips}, {Suntzeff}, {Smith}, {Winkler}, {Winge},
  {Hamuy}, {Hunter}, {Roth}, {Blades}, {Branch}, {Chevalier}, {Fransson},
  {Panagia}, {Wagoner}, {Wheeler}, \& {Harkness}}]{Kirshner_etal_1993}
{Kirshner}, R.~P., {Jeffery}, D.~J., {Leibundgut}, B., {et~al.} 1993, ApJ, 415,
  589

\bibitem[{{Krisciunas} {et~al.}(2011){Krisciunas}, {Li}, {Matheson}, {Howell},
  {Stritzinger}, {Aldering}, {Berlind}, {Calkins}, {Challis}, {Chornock},
  {Conley}, {Filippenko}, {Ganeshalingam}, {Germany}, {Gonz{\'a}lez},
  {Gooding}, {Hsiao}, {Kasen}, {Kirshner}, {Howie Marion}, {Muena}, {Nugent},
  {Phelps}, {Phillips}, {Qiu}, {Quimby}, {Rines}, {Silverman}, {Suntzeff},
  {Thomas}, \& {Wang}}]{Krisciunas_etal_2011}
{Krisciunas}, K., {Li}, W., {Matheson}, T., {et~al.} 2011, \aj, 142, 74

\bibitem[{{Kromer} {et~al.}(2010){Kromer}, {Sim}, {Fink}, {R{\"o}pke},
  {Seitenzahl}, \& {Hillebrandt}}]{Kromer_etal_2010}
{Kromer}, M., {Sim}, S.~A., {Fink}, M., {et~al.} 2010, \apj, 719, 1067

\bibitem[{{Kudelina} {et~al.}(2011){Kudelina}, {Gorbovskoy}, {Balanutsa},
  {Lipunov}, {Kornilov}, {Belinski}, {Shatskiy}, {Tyurina}, {Kuvshinov},
  {Chazov}, {Kuznetsov}, {Zimnukhov}, {Kornilov}, {Tlatov}, {Parhomenko},
  {Dormidontov}, {Yurkov}, {Sergienko}, {Varda}, {Krushinski}, {Zalozhnih},
  {Popov}, {Ivanov}, {Yazev}, {Budnev}, {Konstantinov}, {Chuvalaev},
  {Poleschuk}, {Gres}, {Shumkov}, \& {Shurpakov}}]{Kudelina_etal_2011}
{Kudelina}, I., {Gorbovskoy}, E., {Balanutsa}, P., {et~al.} 2011, The
  Astronomer's Telegram, 3164, 1

\bibitem[{{Li} {et~al.}(2011){Li}, {Bloom}, {Podsiadlowski}, {Miller}, {Cenko},
  {Jha}, {Sullivan}, {Howell}, {Nugent}, {Butler}, {Ofek}, {Kasliwal},
  {Richards}, {Stockton}, {Shih}, {Bildsten}, {Shara}, {Bibby}, {Filippenko},
  {Ganeshalingam}, {Silverman}, {Kulkarni}, {Law}, {Poznanski}, {Quimby},
  {McCully}, {Patel}, {Maguire}, \& {Shen}}]{Li_etal_2011}
{Li}, W., {Bloom}, J.~S., {Podsiadlowski}, P., {et~al.} 2011, \nat, 480, 348

\bibitem[{{Maeda} \& {Iwamoto}(2009)}]{Maeda_Iwamoto_2009}
{Maeda}, K., \& {Iwamoto}, K. 2009, \mnras, 394, 239

\bibitem[{{Maeda} {et~al.}(2009){Maeda}, {Kawabata}, {Li}, {Tanaka}, {Mazzali},
  {Hattori}, {Nomoto}, \& {Filippenko}}]{Maeda_etal_2009}
{Maeda}, K., {Kawabata}, K., {Li}, W., {et~al.} 2009, \apj, 690, 1745

\bibitem[{{Maguire} {et~al.}(2013){Maguire}, {Sullivan}, {Patat}, {Gal-Yam},
  {Hook}, {Dhawan}, {Howell}, {Mazzali}, {Nugent}, {Pan}, {Podsiadlowski},
  {Simon}, {Sternberg}, {Valenti}, {Baltay}, {Bersier}, {Blagorodnova}, {Chen},
  {Ellman}, {Feindt}, {F{\"o}rster}, {Fraser}, {Gonz{\'a}lez-Gait{\'a}n},
  {Graham}, {Guti{\'e}rrez}, {Hachinger}, {Hadjiyska}, {Inserra}, {Knapic},
  {Laher}, {Leloudas}, {Margheim}, {McKinnon}, {Molinaro}, {Morrell}, {Ofek},
  {Rabinowitz}, {Rest}, {Sand}, {Smareglia}, {Smartt}, {Taddia}, {Walker},
  {Walton}, \& {Young}}]{Maguire_etal_2013}
{Maguire}, K., {Sullivan}, M., {Patat}, F., {et~al.} 2013, \mnras, 436, 222

\bibitem[{{Margutti} {et~al.}(2012){Margutti}, {Soderberg}, {Chomiuk},
  {Chevalier}, {Hurley}, {Milisavljevic}, {Foley}, {Hughes}, {Slane},
  {Fransson}, {Moe}, {Barthelmy}, {Boynton}, {Briggs}, {Connaughton}, {Costa},
  {Cummings}, {Del Monte}, {Enos}, {Fellows}, {Feroci}, {Fukazawa}, {Gehrels},
  {Goldsten}, {Golovin}, {Hanabata}, {Harshman}, {Krimm}, {Litvak},
  {Makishima}, {Marisaldi}, {Mitrofanov}, {Murakami}, {Ohno}, {Palmer},
  {Sanin}, {Starr}, {Svinkin}, {Takahashi}, {Tashiro}, {Terada}, \&
  {Yamaoka}}]{Margutti_etal_2012}
{Margutti}, R., {Soderberg}, A.~M., {Chomiuk}, L., {et~al.} 2012, \apj, 751,
  134

\bibitem[{{Marion} {et~al.}(2013){Marion}, {Vinko}, {Wheeler}, {Foley},
  {Hsiao}, {Brown}, {Challis}, {Filippenko}, {Garnavich}, {Kirshner},
  {Landsman}, {Parrent}, {Pritchard}, {Roming}, {Silverman}, \&
  {Wang}}]{Marion_etal_2013}
{Marion}, G.~H., {Vinko}, J., {Wheeler}, J.~C., {et~al.} 2013, \apj, 777, 40

\bibitem[{{Marion} {et~al.}(2009){Marion}, {Garnavich}, {Challis}, {Calkins},
  \& {Peters}}]{Marion_etal_2009dc}
{Marion}, H., {Garnavich}, P., {Challis}, P., {Calkins}, M., \& {Peters}, W.
  2009, Central Bureau Electronic Telegrams, 1776, 1

\bibitem[{{Mazzali} {et~al.}(2013){Mazzali}, {Sullivan}, {Hachinger}, {Ellis},
  {Nugent}, {Howell}, {Gal-Yam}, {Maguire}, {Cooke}, \&
  {Thomas}}]{Mazzali_etal_2013}
{Mazzali}, P., {Sullivan}, M., {Hachinger}, S., {et~al.} 2013, ArXiv e-prints

\bibitem[{{Mazzali} {et~al.}(1997){Mazzali}, {Chugai}, {Turatto}, {Lucy},
  {Danziger}, {Cappellaro}, {della Valle}, \& {Benetti}}]{Mazzali_etal_1997}
{Mazzali}, P.~A., {Chugai}, N., {Turatto}, M., {et~al.} 1997, \mnras, 284, 151

\bibitem[{{Milne} \& {Brown}(2012)}]{Milne_Brown_2012}
{Milne}, P.~A., \& {Brown}, P.~J. 2012, ArXiv e-prints

\bibitem[{{Milne} {et~al.}(2013{\natexlab{a}}){Milne}, {Brown}, {Roming},
  {Bufano}, \& {Gehrels}}]{Milne_etal_2013}
{Milne}, P.~A., {Brown}, P.~J., {Roming}, P.~W.~A., {Bufano}, F., \& {Gehrels},
  N. 2013{\natexlab{a}}, \apj, 779, 23

\bibitem[{{Milne} {et~al.}(2013{\natexlab{b}}){Milne}, {Foley}, \&
  {Brown}}]{Milne_etal_2013_z}
{Milne}, P.~A., {Foley}, R., \& {Brown}, P.~J. 2013{\natexlab{b}}, ApJ
  submitted

\bibitem[{{Milne} {et~al.}(2010){Milne}, {Brown}, {Roming}, {Holland},
  {Immler}, {Filippenko}, {Ganeshalingam}, {Li}, {Stritzinger}, {Phillips},
  {Hicken}, {Kirshner}, {Challis}, {Mazzali}, {Schmidt}, {Bufano}, {Gehrels},
  \& {Vanden Berk}}]{Milne_etal_2010}
{Milne}, P.~A., {Brown}, P.~J., {Roming}, P.~W.~A., {et~al.} 2010, ApJ, 721,
  1627

\bibitem[{{Mould} {et~al.}(2000){Mould}, {Huchra}, {Freedman}, {Kennicutt},
  {Ferrarese}, {Ford}, {Gibson}, {Graham}, {Hughes}, {Illingworth}, {Kelson},
  {Macri}, {Madore}, {Sakai}, {Sebo}, {Silbermann}, \&
  {Stetson}}]{Mould_etal_2000}
{Mould}, J.~R., {Huchra}, J.~P., {Freedman}, W.~L., {et~al.} 2000, ApJ, 529,
  786

\bibitem[{{Nugent} {et~al.}(2011){Nugent}, {Sullivan}, {Cenko}, {Thomas},
  {Kasen}, {Howell}, {Bersier}, {Bloom}, {Kulkarni}, {Kandrashoff},
  {Filippenko}, {Silverman}, {Marcy}, {Howard}, {Isaacson}, {Maguire},
  {Suzuki}, {Tarlton}, {Pan}, {Bildsten}, {Fulton}, {Parrent}, {Sand},
  {Podsiadlowski}, {Bianco}, {Dilday}, {Graham}, {Lyman}, {James}, {Kasliwal},
  {Law}, {Quimby}, {Hook}, {Walker}, {Mazzali}, {Pian}, {Ofek}, {Gal-Yam}, \&
  {Poznanski}}]{Nugent_etal_2011}
{Nugent}, P.~E., {Sullivan}, M., {Cenko}, S.~B., {et~al.} 2011, \nat, 480, 344

\bibitem[{{Parrent} \& {Howell}(2012)}]{Parrent_Howell_2012dn}
{Parrent}, J.~T., \& {Howell}, D.~A. 2012, Central Bureau Electronic Telegrams,
  3174, 2

\bibitem[{{Patat} {et~al.}(2007){Patat}, {Chandra}, {Chevalier}, {Justham},
  {Podsiadlowski}, {Wolf}, {Gal-Yam}, {Pasquini}, {Crawford}, {Mazzali},
  {Pauldrach}, {Nomoto}, {Benetti}, {Cappellaro}, {Elias-Rosa}, {Hillebrandt},
  {Leonard}, {Pastorello}, {Renzini}, {Sabbadin}, {Simon}, \&
  {Turatto}}]{Patat_etal_2007}
{Patat}, F., {Chandra}, P., {Chevalier}, R., {et~al.} 2007, Science, 317, 924

\bibitem[{{Pereira} {et~al.}(2013){Pereira}, {Thomas}, {Aldering}, {Antilogus},
  {Baltay}, {Benitez-Herrera}, {Bongard}, {Buton}, {Canto}, {Cellier-Holzem},
  {Chen}, {Childress}, {Chotard}, {Copin}, {Fakhouri}, {Fink}, {Fouchez},
  {Gangler}, {Guy}, {Hillebrandt}, {Hsiao}, {Kerschhaggl}, {Kowalski},
  {Kromer}, {Nordin}, {Nugent}, {Paech}, {Pain}, {P{\'e}contal}, {Perlmutter},
  {Rabinowitz}, {Rigault}, {Runge}, {Saunders}, {Smadja}, {Tao},
  {Taubenberger}, {Tilquin}, \& {Wu}}]{Pereira_etal_2013}
{Pereira}, R., {Thomas}, R.~C., {Aldering}, G., {et~al.} 2013, \aap, 554, A27

\bibitem[{{Perlmutter} {et~al.}(1999){Perlmutter}, {Aldering}, {Goldhaber},
  {Knop}, {Nugent}, {Castro}, {Deustua}, {Fabbro}, {Goobar}, {Groom}, {Hook},
  {Kim}, {Kim}, {Lee}, {Nunes}, {Pain}, {Pennypacker}, {Quimby}, {Lidman},
  {Ellis}, {Irwin}, {McMahon}, {Ruiz-Lapuente}, {Walton}, {Schaefer}, {Boyle},
  {Filippenko}, {Matheson}, {Fruchter}, {Panagia}, {Newberg}, {Couch}, \& {The
  Supernova Cosmology Project}}]{Perlmutter_etal_1999}
{Perlmutter}, S., {Aldering}, G., {Goldhaber}, G., {et~al.} 1999, ApJ, 517, 565

\bibitem[{{Phillips} {et~al.}(1999){Phillips}, {Lira}, {Suntzeff}, {Schommer},
  {Hamuy}, \& {Maza}}]{Phillips_etal_1999}
{Phillips}, M.~M., {Lira}, P., {Suntzeff}, N.~B., {et~al.} 1999, \aj, 118, 1766

\bibitem[{{Poole} {et~al.}(2008){Poole}, {Breeveld}, {Page}, {Landsman},
  {Holland}, {Roming}, {Kuin}, {Brown}, {Gronwall}, {Hunsberger}, {Koch},
  {Mason}, {Schady}, {vanden Berk}, {Blustin}, {Boyd}, {Broos}, {Carter},
  {Chester}, {Cucchiara}, {Hancock}, {Huckle}, {Immler}, {Ivanushkina},
  {Kennedy}, {Marshall}, {Morgan}, {Pandey}, {de Pasquale}, {Smith}, \&
  {Still}}]{Poole_etal_2008}
{Poole}, T.~S., {Breeveld}, A.~A., {Page}, M.~J., {et~al.} 2008, \mnras, 383,
  627

\bibitem[{{Puckett} {et~al.}(2009){Puckett}, {Moore}, {Newton}, \&
  {Orff}}]{Puckett_etal_2009dc}
{Puckett}, T., {Moore}, R., {Newton}, J., \& {Orff}, T. 2009, Central Bureau
  Electronic Telegrams, 1762, 1

\bibitem[{{Puckett} {et~al.}(2011){Puckett}, {Newton}, {Balam}, {Bohlender},
  {Monin}, {Chisholm}, {Green}, {Gurugubelli}, {Sahu}, {Anupama}, {Anto}, \&
  {Arora}}]{Puckett_etal_2011}
{Puckett}, T., {Newton}, J., {Balam}, D.~D., {et~al.} 2011, Central Bureau
  Electronic Telegrams, 2653, 1

\bibitem[{{Riess} {et~al.}(1996){Riess}, {Press}, \&
  {Kirshner}}]{Riess_etal_1996_mlcs}
{Riess}, A.~G., {Press}, W.~H., \& {Kirshner}, R.~P. 1996, ApJ, 473, 88

\bibitem[{{Riess} {et~al.}(1998){Riess}, {Filippenko}, {Challis},
  {Clocchiatti}, {Diercks}, {Garnavich}, {Gilliland}, {Hogan}, {Jha},
  {Kirshner}, {Leibundgut}, {Phillips}, {Reiss}, {Schmidt}, {Schommer},
  {Smith}, {Spyromilio}, {Stubbs}, {Suntzeff}, \& {Tonry}}]{Riess_etal_1998}
{Riess}, A.~G., {Filippenko}, A.~V., {Challis}, P., {et~al.} 1998, \aj, 116,
  1009

\bibitem[{{Roming} {et~al.}(2005){Roming}, {Kennedy}, {Mason}, {Nousek}, {Ahr},
  {Bingham}, {Broos}, {Carter}, {Hancock}, {Huckle}, {Hunsberger}, {Kawakami},
  {Killough}, {Koch}, {McLelland}, {Smith}, {Smith}, {Soto}, {Boyd},
  {Breeveld}, {Holland}, {Ivanushkina}, {Pryzby}, {Still}, \&
  {Stock}}]{Roming_etal_2005}
{Roming}, P.~W.~A., {Kennedy}, T.~E., {Mason}, K.~O., {et~al.} 2005, Space
  Science Reviews, 120, 95

\bibitem[{{Russell} \& {Immler}(2012)}]{Russell_Immler_2012}
{Russell}, B.~R., \& {Immler}, S. 2012, \apjl, 748, L29

\bibitem[{{Scalzo} {et~al.}(2012){Scalzo}, {Aldering}, {Antilogus}, {Aragon},
  {Bailey}, {Baltay}, {Bongard}, {Buton}, {Canto}, {Cellier-Holzem},
  {Childress}, {Chotard}, {Copin}, {Fakhouri}, {Gangler}, {Guy}, {Hsiao},
  {Kerschhaggl}, {Kowalski}, {Nugent}, {Paech}, {Pain}, {Pecontal}, {Pereira},
  {Perlmutter}, {Rabinowitz}, {Rigault}, {Runge}, {Smadja}, {Tao}, {Thomas},
  {Weaver}, {Wu}, \& {Nearby Supernova Factory}}]{Scalzo_etal_2012}
{Scalzo}, R., {Aldering}, G., {Antilogus}, P., {et~al.} 2012, \apj, 757, 12

\bibitem[{{Scalzo} {et~al.}(2010){Scalzo}, {Aldering}, {Antilogus}, {Aragon},
  {Bailey}, {Baltay}, {Bongard}, {Buton}, {Childress}, {Chotard}, {Copin},
  {Fakhouri}, {Gal-Yam}, {Gangler}, {Hoyer}, {Kasliwal}, {Loken}, {Nugent},
  {Pain}, {P{\'e}contal}, {Pereira}, {Perlmutter}, {Rabinowitz}, {Rau},
  {Rigaudier}, {Runge}, {Smadja}, {Tao}, {Thomas}, {Weaver}, \&
  {Wu}}]{Scalzo_etal_2010}
{Scalzo}, R.~A., {Aldering}, G., {Antilogus}, P., {et~al.} 2010, \apj, 713,
  1073

\bibitem[{{Schaefer} \& {Pagnotta}(2012)}]{Schaefer_Pagnotta_2012}
{Schaefer}, B.~E., \& {Pagnotta}, A. 2012, \nat, 481, 164

\bibitem[{{Schlafly} \& {Finkbeiner}(2011)}]{Schlafly_Finkbeiner_2011}
{Schlafly}, E.~F., \& {Finkbeiner}, D.~P. 2011, \apj, 737, 103

\bibitem[{{Silverman} {et~al.}(2011){Silverman}, {Ganeshalingam}, {Li},
  {Filippenko}, {Miller}, \& {Poznanski}}]{Silverman_etal_2011}
{Silverman}, J.~M., {Ganeshalingam}, M., {Li}, W., {et~al.} 2011, \mnras, 410,
  585

\bibitem[{{Simon} {et~al.}(2009){Simon}, {Gal-Yam}, {Gnat}, {Quimby},
  {Ganeshalingam}, {Silverman}, {Blondin}, {Li}, {Filippenko}, {Wheeler},
  {Kirshner}, {Patat}, {Nugent}, {Foley}, {Vogt}, {Butler}, {Peek},
  {Rosolowsky}, {Herczeg}, {Sauer}, \& {Mazzali}}]{Simon_etal_2009}
{Simon}, J.~D., {Gal-Yam}, A., {Gnat}, O., {et~al.} 2009, \apj, 702, 1157

\bibitem[{{Sternberg} {et~al.}(2011){Sternberg}, {Gal-Yam}, {Simon}, {Leonard},
  {Quimby}, {Phillips}, {Morrell}, {Thompson}, {Ivans}, {Marshall},
  {Filippenko}, {Marcy}, {Bloom}, {Patat}, {Foley}, {Yong}, {Penprase},
  {Beeler}, {Allende Prieto}, \& {Stringfellow}}]{Sternberg_etal_2011}
{Sternberg}, A., {Gal-Yam}, A., {Simon}, J.~D., {et~al.} 2011, Science, 333,
  856

\bibitem[{{Tanaka} {et~al.}(2010){Tanaka}, {Kawabata}, {Yamanaka}, {Maeda},
  {Hattori}, {Aoki}, {Nomoto}, {Iye}, {Sasaki}, {Mazzali}, \&
  {Pian}}]{Tanaka_etal_2010}
{Tanaka}, M., {Kawabata}, K.~S., {Yamanaka}, M., {et~al.} 2010, \apj, 714, 1209

\bibitem[{{Taubenberger} {et~al.}(2011){Taubenberger}, {Benetti}, {Childress},
  {Pakmor}, {Hachinger}, {Mazzali}, {Stanishev}, {Elias-Rosa}, {Agnoletto},
  {Bufano}, {Ergon}, {Harutyunyan}, {Inserra}, {Kankare}, {Kromer},
  {Navasardyan}, {Nicolas}, {Pastorello}, {Prosperi}, {Salgado}, {Sollerman},
  {Stritzinger}, {Turatto}, {Valenti}, \&
  {Hillebrandt}}]{Taubenberger_etal_2011}
{Taubenberger}, S., {Benetti}, S., {Childress}, M., {et~al.} 2011, \mnras, 412,
  2735

\bibitem[{{Taubenberger} {et~al.}(2013){Taubenberger}, {Kromer}, {Hachinger},
  {Mazzali}, {Benetti}, {Nugent}, {Scalzo}, {Pakmor}, {Stanishev},
  {Spyromilio}, {Bufano}, {Sim}, {Leibundgut}, \&
  {Hillebrandt}}]{Taubenberger_etal_2013}
{Taubenberger}, S., {Kromer}, M., {Hachinger}, S., {et~al.} 2013, \mnras, 432,
  3117

\bibitem[{{Theureau} {et~al.}(1998){Theureau}, {Bottinelli}, {Coudreau-Durand},
  {Gouguenheim}, {Hallet}, {Loulergue}, {Paturel}, \&
  {Teerikorpi}}]{Theureau_etal_1998}
{Theureau}, G., {Bottinelli}, L., {Coudreau-Durand}, N., {et~al.} 1998, \aaps,
  130, 333

\bibitem[{{Thomas} {et~al.}(2011){Thomas}, {Aldering}, {Antilogus}, {Aragon},
  {Bailey}, {Baltay}, {Bongard}, {Buton}, {Canto}, {Childress}, {Chotard},
  {Copin}, {Fakhouri}, {Gangler}, {Hsiao}, {Kerschhaggl}, {Kowalski}, {Loken},
  {Nugent}, {Paech}, {Pain}, {Pecontal}, {Pereira}, {Perlmutter}, {Rabinowitz},
  {Rigault}, {Rubin}, {Runge}, {Scalzo}, {Smadja}, {Tao}, {Weaver}, {Wu},
  {(Nearby Supernova Factory)}, {Brown}, \& {Milne}}]{Thomas_etal_2011}
{Thomas}, R.~C., {Aldering}, G., {Antilogus}, P., {et~al.} 2011, ApJ, 743, 27

\bibitem[{{Wang} {et~al.}(2012){Wang}, {Wang}, {Filippenko}, {Baron}, {Kromer},
  {Jack}, {Zhang}, {Aldering}, {Antilogus}, {Arnett}, {Baade}, {Barris},
  {Benetti}, {Bouchet}, {Burrows}, {Canal}, {Cappellaro}, {Carlberg}, {di
  Carlo}, {Challis}, {Crotts}, {Danziger}, {Della Valle}, {Fink}, {Foley},
  {Fransson}, {Gal-Yam}, {Garnavich}, {Gerardy}, {Goldhaber}, {Hamuy},
  {Hillebrandt}, {H{\"o}flich}, {Holland}, {Holz}, {Hughes}, {Jeffery}, {Jha},
  {Kasen}, {Khokhlov}, {Kirshner}, {Knop}, {Kozma}, {Krisciunas}, {Lee},
  {Leibundgut}, {Lentz}, {Leonard}, {Lewin}, {Li}, {Livio}, {Lundqvist},
  {Maoz}, {Matheson}, {Mazzali}, {Meikle}, {Miknaitis}, {Milne}, {Mochnacki},
  {Nomoto}, {Nugent}, {Oran}, {Panagia}, {Perlmutter}, {Phillips}, {Pinto},
  {Poznanski}, {Pritchet}, {Reinecke}, {Riess}, {Ruiz-Lapuente}, {Scalzo},
  {Schlegel}, {Schmidt}, {Siegrist}, {Soderberg}, {Sollerman}, {Sonneborn},
  {Spadafora}, {Spyromilio}, {Sramek}, {Starrfield}, {Strolger}, {Suntzeff},
  {Thomas}, {Tonry}, {Tornambe}, {Truran}, {Turatto}, {Turner}, {Van Dyk},
  {Weiler}, {Wheeler}, {Wood-Vasey}, {Woosley}, \& {Yamaoka}}]{Wang_etal_2012}
{Wang}, X., {Wang}, L., {Filippenko}, A.~V., {et~al.} 2012, \apj, 749, 126

\bibitem[{{Webbink}(1984)}]{Webbink_1984}
{Webbink}, R.~F. 1984, \apj, 277, 355

\bibitem[{{Whelan} \& {Iben}(1973)}]{Whelan_Iben_1973}
{Whelan}, J., \& {Iben}, Jr., I. 1973, \apj, 186, 1007

\bibitem[{{Woosley} \& {Kasen}(2011)}]{Woosley_Kasen_2011}
{Woosley}, S.~E., \& {Kasen}, D. 2011, \apj, 734, 38

\bibitem[{{Woosley} \& {Weaver}(1994)}]{Woosley_Weaver_1994}
{Woosley}, S.~E., \& {Weaver}, T.~A. 1994, \apj, 423, 371

\bibitem[{{Yamanaka} {et~al.}(2009){Yamanaka}, {Kawabata}, {Kinugasa},
  {Tanaka}, {Imada}, {Maeda}, {Nomoto}, {Arai}, {Chiyonobu}, {Fukazawa},
  {Hashimoto}, {Honda}, {Ikejiri}, {Itoh}, {Kamata}, {Kawai}, {Komatsu},
  {Konishi}, {Kuroda}, {Miyamoto}, {Miyazaki}, {Nagae}, {Nakaya}, {Ohsugi},
  {Omodaka}, {Sakai}, {Sasada}, {Suzuki}, {Taguchi}, {Takahashi}, {Tanaka},
  {Uemura}, {Yamashita}, {Yanagisawa}, \& {Yoshida}}]{Yamanaka_etal_2009}
{Yamanaka}, M., {Kawabata}, K.~S., {Kinugasa}, K., {et~al.} 2009, \apjl, 707,
  L118

\bibitem[{{Yaron} \& {Gal-Yam}(2012)}]{Yaron_Gal-Yam_2012}
{Yaron}, O., \& {Gal-Yam}, A. 2012, \pasp, 124, 668

\bibitem[{{Yuan} {et~al.}(2010){Yuan}, {Quimby}, {Wheeler}, {Vink{\'o}},
  {Chatzopoulos}, {Akerlof}, {Kulkarni}, {Miller}, {McKay}, \&
  {Aharonian}}]{Yuan_etal_2010}
{Yuan}, F., {Quimby}, R.~M., {Wheeler}, J.~C., {et~al.} 2010, \apj, 715, 1338

\end{thebibliography}

%% the following is the path to you *.bib file
%% (you do not need to enter the ``.bib'' extention)

%%%%%%%%%%%%%%%%%%%%%%%%%%%%%%%%%%%%%%%%%%%%%%%%%%%%%%%%%%%%%%%%%%%%%%

\end{document}